\documentclass[fleqn,usenatbib]{mnras}
\usepackage[T1]{fontenc}
\usepackage{ae,aecompl}
\usepackage{booktabs}

\usepackage{graphicx}	
\usepackage{amsmath}	
\usepackage{amssymb}	
\usepackage{natbib}
\bibpunct{(}{)}{;}{a}{}{,} 
\usepackage{xcolor}
\newcommand{\es}[1]{}
\usepackage[caption=false]{subfig}

\newcommand{\highlight}[1]{}

\usepackage{etex}	
\usepackage{acronym}  
\usepackage{float}
\usepackage{makecell}
\usepackage{mathtools}
\usepackage{ulem}
\usepackage{footmisc}

\newcommand\myeq{\stackrel{\mathclap{\normalfont\mbox{def}}}{=}}

\acrodef{NS}{neutron star}
\acrodef{SS}{single stripping}
\acrodef{SFR}{star formation rate}
\acrodef{CE}{common-envelope}
\acrodef{MS}{main-sequence}
\acrodef{EC}{electron-capture}
\acrodef{CC}{core-collapse}
\acrodef{US}{ultra-stripped}
\acrodef{IMF}{initial mass function}
\acrodef{BH}{black hole}
\acrodef{BBH}{binary black hole}
\acrodef{BNS}{binary neutron star}
\acrodef{NSBH}{neutron star -- black hole binary}
\acrodefplural{NSBH}[NSBHs]{neutron star -- black hole binaries}
\acrodef{SN}{supernova}
\acrodefplural{SN}[SNe]{supernovae}
\acrodef{BeXRB}{Be X-ray binary}
\acrodefplural{BeXRB}[BeXRBs]{Be X-ray binaries}
\acrodef{HMXRB}{high mass X-ray binary}
\acrodefplural{HMXRB}[HMXRBs]{high mass X-ray binaries}
\acrodef{SMC}{Small Magellanic Cloud}
\acrodef{DCO}{double compact object}
\acrodef{HG}{Hertzsprung gap}
\acrodef{HeHG}{helium Hertzsprung gap}
\usepackage{blindtext}
\usepackage{float}

\title[]{Be X-ray binaries in the SMC as indicators of mass transfer efficiency}

\author[S. Vinciguerra et al.]{
	   Serena Vinciguerra$^{1,2,3}$,
            Coenraad J. Neijssel$^{4,2,3}$,
            Alejandro Vigna-G\'{o}mez$^{6,4,2,3}$,
                      \newauthor  
            Ilya Mandel$^{2,3,4}$,
            Philipp Podsiadlowski$^6$,
            Thomas J. Maccarone$^7$,
            \newauthor
            Matt Nicholl$^{4,8}$,
            Samuel Kingdon$^{4}$,
            Alice Perry$^{4}$,
            Francesco Salemi$^{1}$
\\
$^{1}$ Max Planck Institute for Gravitational Physics (Albert Einstein Institute), D-30167 Hannover, Germany\\
$^{2}$ Monash Centre for Astrophysics, School of Physics and Astronomy, Monash University, Clayton, Victoria 3800, Australia\\
$^{3}$ The ARC Center of Excellence for Gravitational Wave Discovery -- OzGrav\\
$^{4}$ Birmingham Institute for Gravitational Wave Astronomy and School of Physics and Astronomy,\\ University of Birmingham,  B15 2TT, Birmingham, UK\\
$^{5}$ Oxford University, Oxford, OX1 3RH, UK\\
$^{6}$ DARK, Niels Bohr Institute, University of Copenhagen, Blegdamsvej 17, 2100, Copenhagen, Denmark\\
$^{7}$ Department of Physics \& Astronomy, Box 41051, Science Building, Texas Tech University, Lubbock TX 79409-1051, USA \\
$^{8}$ Institute for Astronomy, University of Edinburgh, Royal Observatory, Blackford Hill, EH9 3HJ, UK \\
}

\date{Accepted XXX. Received YYY; in original form ZZZ}
\pubyear{2019}

\begin{document}
\label{firstpage}
\pagerange{\pageref{firstpage}--\pageref{lastpage}}
\maketitle

\begin{abstract}
\acp{BeXRB} consist of rapidly rotating Be stars with neutron star companions accreting from the circumstellar emission disk.  We compare the observed population of \acp{BeXRB} in the Small Magellanic Cloud with simulated populations of \ac{BeXRB}-like systems produced with the COMPAS population synthesis code.  We focus on the apparently higher minimal mass of Be stars in \acp{BeXRB} than in the Be population at large. Assuming that \acp{BeXRB} experienced only dynamically stable mass transfer, their mass distribution suggests that at least $\sim 30\%$ of the mass donated by the progenitor of the neutron star is typically accreted by the B-star companion. We expect these results to affect predictions for the population of double compact object mergers.
A convolution of the simulated \ac{BeXRB} population with the star formation history of the Small Magellanic Cloud shows that 
 the excess of \acp{BeXRB}
 is most likely explained by this galaxy's burst of star formation $\sim$20--40 Myr ago.
\end{abstract}

\section{Introduction}
Be stars are classically defined as \ac{MS} stars of spectral type B  (e.g. \citealt{Rivinius2013, belczynski2009apparent}), although the so called Be phenomenon is recognised to extend from early A to late O spectral types, i.e., masses down to 
$\sim 3\,\mathrm{M_{\odot}}$ (\citealt{Rivinius2013} and references therein).
The Be phenomenon refers to the possibly transient presence of Balmer emission lines  in the spectrum of a non-supergiant star \citep{Rivinius2013}.
Therefore in the Be nomenclature {\it B} stands for the most common spectral type and {\it e} for the Balmer emission lines.
These emission lines trace the presence of a surrounding decretion disk, which is composed of material outflowing from the Be star, and may appear and disappear together with the disk during the star's life.
The decretion disk is strongly linked to high rotational velocities:
Be stars are among the most rapid non-compact rotators,
 with an average velocity of $\gtrsim 75\%$ of the Keplerian limit at the equator \citep{Rivinius2013}.
Common explanations for such high rotational velocities are
initial rotation, evolution during the star's \ac{MS} lifetime toward Keplerian velocity \citep{ekstrom2008evolution}, and interactions with a companion including mass transfer episodes, tidal locking and mergers (e.g. \citealt{DeMink2013}).

\acp{BeXRB} represent a high fraction of \acp{HMXRB}; they are
composed of a compact object and a Be star \citep{Rappaport1982, vanDenHeuvel1987,Reig2011}. 
The compact object accretes from the decretion disk of the Be star. 
With the possible exception of the \ac{BH} in MWC 656 \citep{Casares2014,Munar_Adrover_2014}, only slowly rotating \acp{NS}\footnote{Two other objects have Be stars which interact with compact objects, but do not show the same phenomenological X-ray behavior as the other \acp{BeXRB}.  These are PSR B1259-63, which has a fast radio pulsar, and LSI +61 303, the nature of which is debated.} have been successfully identified as compact objects in \acp{BeXRB} \citep{klus2013spin}.
The accretion onto the \ac{NS} predominantly  occurs close to periastron and is thought to be at the origin of the \ac{BeXRB} outbursts \citep{okazaki2001natural,okazaki2007active}.

These X-ray outbursts are observed from binaries in the Milky Way and the Magellanic Clouds (\citealt{Reig2011} and references therein).
The \ac{SMC} is particularly rich in \acp{BeXRB} for its mass, with $\sim 70$ confirmed systems \citep{Coe&Kirk2015} compared to $\sim 60-80$ in the Galaxy \citep{Reig2011,Walter2015,presentation}; 
indeed, all the classified \acp{HMXRB} in the \ac{SMC} except SMC X-1 \citep{Haberl&Sturm2015} are \acp{BeXRB}. 
As a sub-population of \acp{HMXRB}, \acp{BeXRB} are characterised by orbital periods in the range of tens to hundreds of days.  Be stars in \acp{BeXRB} are found at early spectral types (no later than B5 in the \citealt{Coe&Kirk2015} \ac{SMC} catalogue), suggesting masses $\gtrsim 8\, \mathrm{M_{\odot}}$ \citep{Reig2011, presentation}, significantly higher than for single Be stars.

Here we investigate the origin of \acp{BeXRB}, by comparing 
observations from the \ac{SMC} catalog by \citet{Coe&Kirk2015}
against mock populations of \ac{BeXRB}-like systems simulated with the rapid binary population synthesis code COMPAS \citep{stevenson2017formation,Barrett2018,vigna2018formation,Coen2019}. 
Rapid population synthesis allows us to easily explore the physics of binary evolution.
We focus the comparison on the total number of systems and their mass and orbital period distributions.  

Our study suggests that the number of Be stars in interacting binaries is enhanced by accretion-induced spin-up during dynamically stable mass transfer.  We conclude that $\gtrsim$ 30\% of the mass lost by the \ac{NS}-progenitor donor should typically be accreted by the B-star companion to match the observed mass distribution.  
These findings impact other interacting binary populations, including the formation of double compact objects.

In section \ref{sec:SMC}, we recap the properties and most relevant selection effects impacting the \acp{BeXRB} reported in the \ac{SMC} catalogue of \citet{Coe&Kirk2015}. 
We briefly describe COMPAS and outline the parameters of our simulations in section \ref{sec:COMPAS}. 
We describe the conversion of population synthesis results into \ac{SMC} predictions in section \ref{sec:post_processing}. 
We present our findings in section \ref{sec:Results} and discuss them in section \ref{sec:discussion}. 
We conclude in section \ref{sec:Conclusions}.

\section{Observed sample}
\label{sec:SMC}
We use the 69 \ac{SMC} \acp{BeXRB} listed in the catalogue \citet{Coe&Kirk2015} (CK catalogue hereafter).

The orbital period is reported for 44 of these systems 
(see figures \ref{fig:MMS_main} and \ref{fig:Porb_main_MMS_secondary}).
To estimate the orbital period of the remaining 25 \acp{BeXRB}, we use the Corbet relation between the \ac{NS} spin and orbital periods \citep{Corbet1984}. We fit a linear relationship between the logs of the measured orbital and \ac{NS} spin periods from the CK catalogue,
\begin{equation}\label{eq:Corbet}
\log_{10}\left(\frac{P_\mathrm{orb}}{\mathrm{days}}\right) = 0.4329 \log_{10}\left(\frac{P_\mathrm{spin}}{\mathrm{s}}\right) + 1.043 
\end{equation}
and apply it to the listed \acp{BeXRB} with unknown orbital periods.

The CK catalogue lists the eccentricity of only 7 binaries. 
The eccentricities of \acp{BeXRB} have typically been estimated using pulse timing data, with the Doppler shifts on the pulsations at different orbital phases 
\citep{Coe&Kirk2015,Townsend2011}.
The relatively long period and transient nature of \acp{BeXRB} make eccentricity measurements quite challenging, particularly for the less eccentric orbits (see e.g. upper limits in \citealt{Townsend2011}).

The uncertain eccentricities also prevent accurate dynamical measurements of the Be star masses.
However their spectral type distribution \citep{Antoniou2009,Maravelias2014} suggests that their masses do not extend to the low values observed for the general population of other Be stars (down to $\sim \,3\,\mathrm{M_{\odot}}$). 
While the difference between spectral types of Be stars as a population and Be stars in \acp{BeXRB} is clear, the corresponding difference in Be star masses is not straightforward to estimate.
Mass estimates from spectral and luminosity types are subject to several uncertainties, concerning both the classification procedures and the physics of the systems. 
Phenomena such as dust extinction and rotational mixing may bias the inferred properties of the star, and thus its spectral and luminosity classifications.
Moreover, the dependence of the star's spectrum and luminosity on its mass is also partly degenerate with its age and chemical composition.
As a conservative estimate for the minimum mass of \acp{BeXRB} in the CK catalogue, we use the minimum value of $\mathrm{6\,M_\odot}$ reported in table 4 of \citet{Hohle2010}.

In the following, we qualitatively present the main observational selection biases which likely affect the observed population of \acp{BeXRB}.

\subsection{A qualitative understanding of the selection biases}
\label{sec:selection_effects}
The bulk of the \ac{SMC} \ac{BeXRB} \acp{NS} reported in the CK catalogue were found in {\it RXTE} scans.
These scans typically monitored the \ac{SMC} on a roughly weekly basis for over a decade, with typical exposure times of $10^4$ seconds, but with continuous observations limited to about 3000 seconds due to Earth occultations that occur on the satellite orbital period of 96 minutes.  Orbital periods have been found through a mixture of pulse timing (e.g. \citealt{Townsend2011}), photometric periodicities \citep{Schurch2011}, and periodicities of the outbursts from the eccentric binaries' periastron passages.  

We note that there is a substantial bias against both detecting wide systems and measuring their orbital periods. 
Some of the key issues concerning this problem are discussed in \citet{Laycock_etal2010}: the wider systems should be fainter and their periastron passages less frequent.  Beyond that, in the pulsation searches, the long pulse period systems are, again, much less likely to be discovered. 
All the systems with spin periods longer than 500 seconds are indeed discovered by imaging satellites like {\it Chandra} and {\it XMM-Newton} whose \ac{SMC} sampling is considerably poorer than that of {\it RXTE} \citep{Laycock_etal2010}. 
Given that \acp{BeXRB} show a strong correlation between orbital period and spin period \citep{Corbet1984}, this, in turn, leads to a bias against the long orbital period systems. 
We can see this by splitting the sample of 63 \ac{BeXRB} of \citet{Haberl&Sturm2015} in two, according to their \ac{NS} spin period. Of the 32 binaries with \ac{NS} spin periods equal to or below the median, 29 have measured orbital periods; this is true only for 19 systems of the 31 binaries with \ac{NS} spin periods above the median.  If the orbital period measurement had a uniform success rate of 29/32, the binomial probability of observing only 19 of 31 orbital periods in the long spin period group would be 1.1$\times10^{-5}$.

Such a bias is to be expected, for a variety of reasons.  At long periods, the rates of change of spin frequency due to accretion torques typically seen during outbursts (e.g., \citealt{1997ApJS..113..367B}) are $\sim1-2$ orders of magnitude larger  than those due to Doppler shifts, preventing orbital period measurements via Doppler shifts in pulse timing.  Optical spectroscopic monitoring of very long period binaries has typically not been practical due to the challenges in scheduling such observations.  Photometric period estimation is notoriously susceptible to incorrect identification due to aperiodic noise (e.g. \citet{Press1978ComAp}), and furthermore, the longer period systems are likely to have weaker photometric modulations on the orbital period.  We hope that in the new era of large numbers of queue-scheduled telescopes, an efficient management of the observing time will help fill in the missing orbital periods.

A further penalty impacting the widest \acp{BeXRB} concerns the Be decretion disks, which in these cases may never reach the Roche lobe of the \ac{NS}, or may do so only intermittently, near periastron passages (as is the case for the Galactic system PSR B1259-63, which is a gamma-ray binary near periastron and a radio pulsar binary near apastron -- see e.g. \citealt{Chernyakova14}), significantly limiting the X-ray emission. 
Pulsar searches with the Square Kilometer Array likely represent the best path forward to discovering these systems, but in some cases, the Be star lifetimes may exceed the lifetimes of the systems as active pulsars, and in other cases, the pulsar's opening angle may not be pointed toward Earth, so it is likely that only a statistical sampling of these objects will be obtained with instruments available in the next few decades. 
Moreover, according to \citet{Reig2011}, a significant fraction of the widest Galactic systems are persistent sources, generally characterised by weaker X-ray emission.
Similar statistics in the \ac{SMC} could bias the observed sample of \acp{BeXRB} against long-orbital-period binaries.
Finally, instabilities in the Be decretion disks can also impact the long-term detectability of \acp{BeXRB} \citep{Rivinius2013}.

\section{Population synthesis code: COMPAS}
\label{sec:COMPAS}
To study the evolution of massive stellar binaries we use the population-synthesis suite of COMPAS (\url{http://compas.science}). 
By rapidly evolving large populations of binaries we can perform statistical studies on the physics of stellar and binary evolution. 
Similarly to other rapid population-synthesis codes (e.g. \citealt{Zwart1996, Nelemans2001, hurley2002evolution,Izzard2004,Belczynski2008,Toonen2012, Giacobbo2018, 2019Breivik}),
we rely on analytic approximations of a set of pre-calculated models of single stars and stellar winds to reduce the computational cost (\citealt{hurley2000comprehensive, stevenson2017formation} and references therein).

We Monte Carlo sample the initial parameters of the binaries.  The primary star, which we define to be the initially more massive component of the binary, follows the \ac{IMF} of \citet{2001MNRAS.322..231K}, with a minimum mass of $5~\mathrm{M_{\odot}}$ and a maximum mass of $150~\mathrm{M_{\odot}}$. We draw the secondary (hereafter the lighter companion at zero-age \ac{MS}) from a flat mass ratio distribution in the range $q\in[0,1]$ \citep{2012SanaInteraction}. 
For the binary separation $a$, we assume a uniform probability density in log $a$
\citep{opik1924statistical}, although \citet{MoeDiStefano:2017} more recently suggested coupled initial distributions. 
We set the metallicity to $Z = 3.5\times 10^{-3}\sim Z_{\mathrm{SMC}}$ \citep{Davies2015}. 
For each simulation, we evolve a population of $3\times 10^5$ binaries in order to obtain a statistically accurate representation of the population of interest.

We do not attempt to follow the evolution of stellar rotation rates within COMPAS, given many uncertainties regarding: spin-up during mass transfer, spin-down through winds, and angular momentum transport within stars or, especially, through the decretion disk.  Instead, we compare the observed samples of \acp{BeXRB} in the \ac{SMC} against all \ac{NS}+\ac{MS} binaries in COMPAS models, without labelling stars as B or Be (but see, e.g., \citealt{belczynski2009apparent,grudzinska2015formation} for a different approach).  
We qualitatively discuss the impact of processes affecting spin evolution, including mass loss and tidal effects, in section \ref{sec:discussion}. 

We generally follow the stellar and binary evolution prescription of \citet{vigna2018formation}; we summarise the key assumptions and highlight exceptions below.  
Two particularly important sets of choices for \acp{BeXRB} are those related to supernovae and mass transfer. In the following we present our default assumptions and some relevant variations, which we use to compare against observations.

\subsection{Supernovae}
\label{sec:SN}
We determine the properties of the compact object remnant according to the delayed prescription of \citet{2012FryerRemnants}. 
The remnant gets a kick from the asymmetric ejection of material. The magnitude of the kick depends on the type of \ac{SN} and the mass of the pre-explosion core, including the amount of stripping during the previous mass transfer episodes (\citealt{vigna2018formation} and references therein).
We follow \citet{vigna2018formation} in drawing the natal kicks from a Maxwellian distribution, whose 1D root mean squared velocity is set to $\sigma_{\mathrm{1D}} = 265\,\mathrm{km/s}$ for \ac{CC} \acp{SN} and $\sigma_{\mathrm{1D}} = 30\, \mathrm{km/s}$ for \ac{EC} and \ac{US} \acp{SN} \citep{Lyne1994, Hansen1997, Cordes1998, Arzoumanian2002, Pfahl2002,Podsiadlowski2004,Hobbs:2005,Schwab2010, Suwa:2015saa,Tauris2015, Janka2017,Gessner:2018ekd,Muller:2018utr,2020Powell}. 
We then re-scale the drawn velocity by a ``fallback factor'', which depends on the CO core mass of the progenitors.
This factor effectively introduces a difference between \ac{NS} and \ac{BH} natal kick velocities, with reduced kicks for the latter.

When a star, already stripped of its hydrogen envelope, overflows its Roche lobe, it initiates a further episode of mass transfer. If this interaction removes the entire helium envelope from the donor, the star may experience a \ac{SN} with reduced natal kicks, an \ac{US}\ac{SN} \citep{Pfahl2002, Podsiadlowski2004, Tauris2015}. 
One motivation for this low-kick assumption is that this second episode of mass transfer generally leads to a 
low core binding energy; this in turn allows for a rapid \ac{SN} explosion, during which aspherical  instabilities that may be responsible for the \ac{SN} kicks do not have time to develop \citep{Podsiadlowski2004}. 
Our implementation of \ac{US}\acp{SN} differs from the one adopted in \citet{vigna2018formation}, where reduced natal kicks are only applied if the \ac{NS} progenitors has lost its helium envelope while interacting with a \ac{NS}. 
\ac{US}\acp{SN} and mass transfer initiated by stripped stars are active research topics, further investigations are therefore required to determine 
the conditions for the validity of our 
assumptions (e.g. the stability of such mass transfer episodes and the removal of the entire helium envelope \citealt{Tauris2015,Laplace2020}). 

Unlike \citet{vigna2018formation} and \citet{hurley2000comprehensive}, our binary evolution model assumes that a star undergoes an \ac{EC}\ac{SN} if the mass of its core at the base of the asymptotic giant branch ranges between $1.83~\mathrm{M_\odot}$ and $2.25~\mathrm{M_\odot}$ \citep{2012FryerRemnants}. 
We follow \citet{2012FryerRemnants} in setting the maximum mass of a \ac{NS} to $2.5\,\mathrm{M_\odot}$.
Moreover, compared to previous COMPAS versions, we are now allowing stars with carbon-oxygen core masses at \acp{SN} above $1.38~\mathrm{M_\odot}$ to collapse into \acp{NS} or \acp{BH}. 

\subsection{Mass transfer}
\label{subsection:masstransfer}
Mass transfer crucially influences the orbital period and component masses. 
Mass transfer starts when a star overflows its Roche lobe.  We determine the dynamical stability of mass transfer by comparing the radial response of the donor star and the response of the Roche lobe radius to mass transfer (see section 2.2.4 of \citealt{vigna2018formation} and section \ref{sec:stability-method} below).

When the mass transfer is dynamically unstable, we assume that the system experiences a \ac{CE} event (for a review see \citealt{ivanova2013common}). 
The core of the donor and the accretor orbit inside the donor's former envelope, which is no longer co-rotating with the binary.  We adopt the $\alpha-\lambda$ parameterisation, based on the description proposed by \citet{Webbink1984} and \citet{Kool1990}, to describe the outcome of a \ac{CE} event.  In this formalism, $\lambda$ parametrises the binding energy of the envelope and $\alpha$ parametrises the efficiency with which orbital energy can be used to expel the envelope. 
We follow \citet{vigna2018formation} in assuming that $\alpha=1$.
We use fits by \citet{2010XuLambda,XuLi2010erratum}, as implemented in StarTrack \citep{Dominik2012}, for the binding energy parameter $\lambda$. 
Furthermore we assume that the companion star does not accrete during the \ac{CE} phase. 

\subsubsection{Accretion efficiency}
\label{sec:beta}
According to our model, stable mass transfer proceeds on a nuclear or thermal timescale, depending on the evolutionary phase of the donor.  The companion may therefore accrete a significant amount of mass during this process.
We denote the ratio of accreted mass $\Delta M_{\mathrm{acc}}$ to mass lost by the donor $\Delta M_{\mathrm{donor}} $ with the efficiency parameter $\beta$:
\begin{equation}
\beta\, \myeq\, \Delta M_{\mathrm{acc}}/\Delta M_{\mathrm{donor}}  \, .
\label{eq:beta}
\end{equation}

Our default prescription for the accretion efficiency of mass transfer prior to the current study, motivated by \citet{hurley2002evolution}, estimates the accretion efficiency $\beta$ by comparing the mass loss rate of the donor $\dot{M}_\mathrm{donor}$ to the maximal mass acceptance rate of the accretor $\dot{M}_\mathrm{acc,max}$ 
\begin{equation}
\beta = \min\left(1, 10 \times \frac{\dot{M}_\mathrm{acc,max}}{\dot{M}_\mathrm{donor}}\right).
\label{eq:betaTH}
\end{equation}
The steady-state mass acceptance rate is set by the time required for the accretor to radiate away the energy carried by the in-falling matter, and is therefore inversely proportional to the accretor's thermal or Kelvin-Helmholtz timescale:
\begin{equation}
\dot{M}_\mathrm{acc,max} \sim \left(\frac{\varepsilon_\mathrm{g,acc}}{L_\mathrm{acc}}\right)^{-1}\propto R_\mathrm{acc},
\label{eq:th}
\end{equation}
where $\varepsilon_\mathrm{g,acc}$ is the specific gravitational binding energy at the accretor's surface, $L_\mathrm{acc}$ is the accretor's luminosity and $R_\mathrm{acc}$ is its radius at the beginning of mass transfer.
In equation \ref{eq:betaTH} we include a factor of ten \citep{Paczynski1972,hurley2002evolution} to approximately account for the change in the accretor's thermal timescale due to mass transfer, including its expansion and increased luminosity. 
Hereafter we label this prescription for the accretion efficiency parameter $\beta$ as $\mathrm{THERMAL}$.

Alternatively, we can account for the expansion of the accretor up to the point of filling its Roche lobe (assuming the mass transfer to be fully conservative until this point) by using the accretor's actual Roche lobe radius in place of $R_\mathrm{acc}$ in equation (\ref{eq:th}) and dispensing with the factor of $10$ in equation (\ref{eq:betaTH}).  We observe no significant difference between these two approaches for \ac{BeXRB} predictions. In the last simplistic model we do not explicitly account for the change in the accretor luminosity during mass transfer (see e.g. \citealt{KippenhahnMeyerHofmeister:1977}).

In the following, we explore the impact of accretion efficiency by additionally testing different fixed $\beta$ values. 
Despite the lack of physical foundation, these variations 
can help us understand the impact of accretion efficiency during mass transfer episodes preceding the \ac{BeXRB} phase. 
We vary the accretion efficiency of mass transfer by setting $\beta$ to 0.5, 0.75, or 1. 
When these fixed $\beta$ values are applied, mass transfer episodes initiated by \ac{MS} donors and post-helium core burning stripped stars are, however, always assumed fully conservative
\footnote{Stable mass transfer initiated by \ac{MS} stars is commonly accepted to be close to conservative \citep{Schneider2015}; it is similarly so also in our default model (see equation \ref{eq:betaTH}). The accretion efficiency during mass transfer episodes initiated by evolved stripped stars is less studied; we fix its value to 1 to be conservative in our conclusions (see section \ref{sec:Results}).}.

If the accretor is a compact object, we assume that the accretion is Eddington-limited, regardless of the $\beta$ model assumed for other mass transfer episodes. 

\subsubsection{Specific angular momentum loss}
\label{sec:gamma}
When $\beta <1$, i.e. the mass transfer is non-conservative, mass is lost from the binary system, taking away some orbital angular momentum. 
In COMPAS, this angular momentum loss is parametrised by 
\begin{equation}
\gamma\, \myeq\, h_{\mathrm{loss}}/h_{\mathrm{binary}} \, ,
\label{eq:gamma}
\end{equation}
the ratio between the specific angular momenta of the ejected material $h_{\mathrm{loss}}$ and the binary $h_{\mathrm{binary}}$. The specific angular momentum (total angular momentum divided by total mass) of a circular binary with mass ratio $q$, total mass $M$ and separation $a$ is $h_{\mathrm{binary}} = q(1+q)^{-2}\sqrt{MGa}$, where $G$ is the universal gravitational constant and we defined the mass ratio $q$ as $q = M_{\mathrm{acc}}/M_{\mathrm{donor}}$.

In our default model, we assume that the mass lost from the binary during mass transfer  instantaneously takes away the angular momentum it had at the surface of the accretor \citep{hurley2002evolution, stevenson2017formation}.
This mode of mass loss is commonly referred to as isotropic re-emission ($\mathrm{ISO}$ hereafter in equations and figures) and
 corresponds to $\gamma_{\mathrm{ISO}} = q^{-1}$.

\begin{figure}
\begin{centering}
\includegraphics[width=8.75cm] {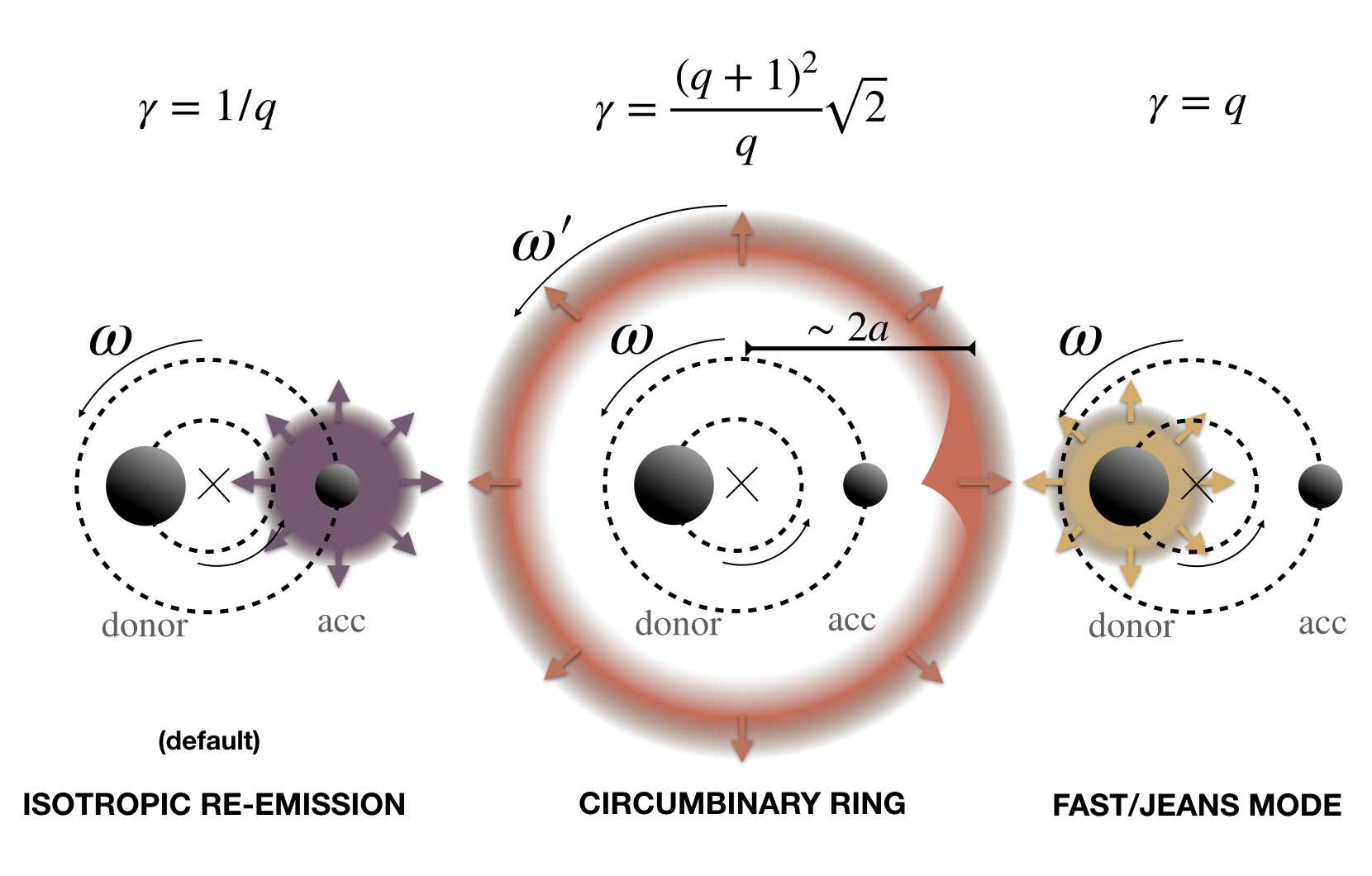}
\caption{A schematic representation of the most commonly used angular momentum loss models and their underlying assumptions in terms of where mass is lost from. In this plot, $\mathrm{acc}$ is an abbreviation for accretor. The colours of the matter leaving the systems for each specific angular momentum prescription match the colours used to represent results in figure \ref{fig:Porb_main_MMS_secondary}.
}
\label{fig:gammas}
\end{centering}
\end{figure}

Other common scenarios consider matter leaving from:
\begin{itemize}
\item the surface of the donor, fast or Jeans mode ($\mathrm{JEANS}$ heareafter), according to which $\gamma_{\mathrm{JEANS}} = q$; 
\item or a circumbinary ring ($\mathrm{CIRC}$ heareafter).
 In our settings, the semimajor axis of the circumbinary ring $a_{\mathrm{ring}}$ 
is fixed to twice the binary's semimajor axis $a$ \citep{Artymowicz1994}, so that the specific angular momentum of the mass lost by the system is $\gamma_{\mathrm{CIRC}}$ times the specific angular momentum of the binary, with:
\begin{equation}
  \gamma_{\mathrm{CIRC}} = \displaystyle\frac{(1+q)^2}{q}\sqrt{\frac{a_{\mathrm{ring}}}{a}} = \displaystyle\sqrt{2}\frac{(1+q)^2}{q} \, .
 \label{eq:gamma_circ}
\end{equation}
\end{itemize}
 In figure \ref{fig:gammas}, we show a schematic representation of the three different scenarios assumed for the angular momentum loss during mass transfer. From left to right, we show the case of matter leaving the binary from: the surface of the accretor (as in our default and, as shown below, preferred models); a circumbinary ring with semimajor axis twice that of the binary; and the surface of the donor. The colours match those used in figure \ref{fig:Porb_main_MMS_secondary}.

\subsubsection{Stability criteria}
\label{sec:stability-method}
In COMPAS, the stability criteria are based on the stellar type of the donor and on the response of the Roche lobe to mass loss, which in turn depends on the values of $\beta$ and $\gamma$ of the specific system. 
The stellar type of the donors is an indicator for the star's response to adiabatic mass loss.
This response can be parametrised by $\zeta^*$, which is defined in terms of the star's radius $R^*$ and its mass $m$ as:
\begin{equation}
    \zeta^* \equiv \displaystyle\frac{d\ln(R^*)}{d\ln(m)}\, .
\end{equation}
For early phases of the donor's evolution, we adopt fixed values of $\zeta^*$, based on the results of \citet{Ge2015}; specifically, we use $\zeta^*_{MS} = 2.0$ for \ac{MS} donors and  $\zeta^*_{HG}  = 6.5$ for \ac{HG} donors (hydrogen shell burning with contracting helium core). For later evolutionary stages of a star with a hydrogen envelope, we use fits to condensed polytrope models \citep{Soberman1997}.
$\zeta^*$ is then compared to $\zeta_{\mathrm{RL}}(\beta,\gamma)$, which represents the response of the radius of the Roche lobe to mass transfer. 
The mass transfer episode is considered dynamically stable if $\zeta^*\geq \zeta_{\mathrm{RL}}$.
Mass transfer initiated after the end of core helium burning of a stripped donor is assumed to always be stable \citep{vigna2018formation}.

As we show below, our models suggest that mass transfer episodes initiated by \ac{HG} stars are particularly important for the formation of \acp{BeXRB}. 
In our default model, our assumptions determine a critical mass ratio of $\sim 0.23$, compared e.g. to the \citet{Claeys2014} value of $0.25$\footnote{COMPAS and the analysis in \citet{Claeys2014} share some assumptions, such as stellar evolution fits \citep{hurley2000comprehensive} to the tracks of \citet{Pols1998}.}:
only systems with mass ratio above this value can experience stable mass transfer initiated by \ac{HG} stars. 
In table \ref{tab:qcrit} we report the critical mass ratio for mass transfer episodes with \ac{HG} donors, assuming different models for the accretion efficiency $\beta$ and the specific angular momentum loss $\gamma$.
\begin{table}
\centering
\begin{tabular}{|l|c|c|c|}\hline
\diaghead{\theadfont JEANSSsss}%
  {$\gamma$}{$\beta_{\mathrm{HG}}$}&\thead{0}&\thead{0.5}&\thead{1}\\
\hline
{\it ISO} & $ \sim 0.22 $ & $ \sim 0.24 $& $\sim 0.26$\\
\hline
{\it CIRC}  & $ \sim 0.59$ &$  \sim 0.38$ & $\sim 0.26$\\
\hline
{\it JEANS} & all stable & $\sim  0.14 $ & $\sim 0.26$\\
\hline 
\end{tabular}
\caption{Critical mass ratios for mass transfer initiated by \ac{HG} stars, depending on the assumed  accretion efficiency $\beta$ and specific angular momentum loss $\gamma$. 
In our model, the stability threshold is determined for each binary by comparing the responses of the donor radius and the Roche lobe size to mass transfer. According to our default model, the donor response to mass loss depends on its stellar type and is fixed to a logarithmic derivative of 6.5 for a $\mathrm{HG}$ star. The response of the Roche lobe to mass loss is determined by the specific angular momentum lost, the accretion efficiency and the mass ratio of each binary, so the table lists typical values. In the table we show that, assuming isotropic re-emission, the critical mass ratio can vary between $0.22$ and $0.26$ depending on the choice of $\beta$. These values can be compared with the fixed mass ratio of $0.25$ adopted in \citep{Claeys2014}, who also assumed isotropic re-emission from the surface of the accretor.
}
\label{tab:qcrit}
\end{table}

We also explore an alternative stability criterion, this time based on the temperature of the donor, instead of its stellar type. 
In this model, when a post-\ac{MS} star fills its Roche lobe, we approximate the response of the stellar radius to mass loss with a constant logarithmic derivative for radiative envelopes, if its temperature is above $\log_{\mathrm{10}}(T/K) = {3.73}$ (i.e., $T\gtrsim5370$~K), following \citet{Belczynski2008}. For cooler surface temperatures we use condensed polytrope models appropriate for stars with deep convective envelopes \citep{Soberman1997}. 
The stability of mass transfer is then again determined by comparing the stellar response with the Roche lobe response. 
In contrast to our default model, this variation for determining the stellar response allows early core helium burning stars to engage in stable mass transfer for a broader range of mass ratios, and so contribute to the formation of \ac{NS}+\ac{MS} systems.

\section{Modelling the \ac{SMC} population}
\label{sec:post_processing}

To compare our results with observations, we convert COMPAS population synthesis results into a simulated \ac{SMC} population of \ac{NS}+\ac{MS} binaries. 
The distribution of orbital periods $P_\mathrm{orb}$ and companion masses $M_\mathrm{MS}$ of current \ac{NS}+\ac{MS} systems is given by
\begin{equation}
\frac{d N}{d P_\mathrm{orb} d M_\mathrm{MS}} = \int_{-\infty}^{0} \frac{d M_\mathrm{SF}}{d t} (-t) 
\frac{d N}{d M_\mathrm{SF} d P_\mathrm{orb} d M_\mathrm{MS}} (t) d t\, .
\label{eq:integral}
\end{equation}
The first term in the integrand is the \ac{SMC} star formation rate history (measured at time $t$ {\it before} the observing time 0).  We use the \ac{SMC} \ac{SFR} history of \citet{rubele2015vmc} for this convolution.  The second term is the number of \ac{NS}+\ac{MS} binaries per unit orbital period per unit companion mass per unit star-forming mass at time $t$ after star formation, as simulated by COMPAS.  It can be viewed as the efficiency function of converting a mass $M_\mathrm{SF}$ of star formation into a number density of  \ac{NS}+\ac{MS} systems per unit orbital period per unit mass of the \ac{MS} star, for systems still present after a time $t$ since their formation.  

In practice, this integral is evaluated through a Monte Carlo approximation using our synthetic populations.  
The minimal delay time from star formation to observation is 
set by the time required for each binary of interest to evolve from zero-age \ac{MS} to the first \ac{SN}. 

We estimate statistical uncertainties from this integration to be of the same order as the ones due to the imperfectly known \ac{SMC} \ac{SFR} history as shown in figure 16 of \citet{rubele2015vmc}.  
Both sources of uncertainty are combined to define the one-$\sigma$ prediction range, indicated with shaded areas in figures \ref{fig:MMS_main}, \ref{fig:Porb_main_MMS_secondary}, \ref{fig:Temperature_US_SS} and \ref{fig:stability}.

\section{Results}
\label{sec:Results}
In this section we present our findings. We outline the most likely formation channel in \ref{sec:formationCH} and compare observations with the properties of the simulated \acp{BeXRB} in \ref{sec:comparisons}.

 We distinguish between: (i) the {\it simulated intrinsic} population, which comprises the binary samples directly produced by COMPAS evolutionary models and (ii) the {\it simulated \ac{SMC}} population, which is obtained by accounting for the \ac{SFR} of the \ac{SMC} \citep{rubele2015vmc} and the duration of the \ac{NS}+\ac{MS} phase, set in our analysis by the duration of the \ac{MS} lifetime of the secondary (see section \ref{sec:post_processing}). Figures and tables refer to the simulated \ac{SMC} population, which is also assumed throughout the text, unless otherwise specified.

\subsection{Formation channel}
\label{sec:formationCH}
We compare observations of \acp{BeXRB} in the \ac{SMC} against the predicted properties of \ac{BeXRB}-like systems in the synthetic populations of binaries evolved with COMPAS.
Unlike some other stellar population synthesis studies (e.g. \citealt{belczynski2009apparent,grudzinska2015formation}), 
we avoid {\it a priori} cuts on orbital periods based on \ac{BeXRB} observations.
We instead investigate the origin of the observed systems by exploring the key predictions for different formation channels, expanding on the approach proposed for general Be stars by \citet{presentation}.
In this subsection we present the characteristics of the simulated systems containing an intermediate or high mass 
\ac{MS} star orbiting a \ac{NS}. We do not {\it a priori} assume any prevalent formation channel for Be stars in \acp{BeXRB}.

According to our simulations, systems that avoid mass transfer can only marginally contribute to the observed sample of \acp{BeXRB}.
Indeed the orbital periods which characterise non-interacting binaries are typically much longer than observations ($\gtrsim10$ years).
The absence of very long orbital period systems in observations is entirely consistent with selection effects (see section  \ref{sec:selection_effects}).  Therefore, observations do not constrain the possibility of forming Be stars in \ac{BeXRB} systems by means of initial rotation or evolution toward critical velocity in the absence of interactions with a binary companion.

Our simulations also suggest that binaries whose first mass transfer episode was dynamically unstable do not match the observed companion masses in \ac{BeXRB}.  
According to our evolutionary model, dynamical instability is favoured for high mass ratio systems; for the case of \acp{BeXRB}, this implies low-mass \ac{MS} accretors.
Moreover we expect minimal accretion during the short-duration \ac{CE} phase.
 Therefore, in most \ac{NS} + \ac{MS} star systems that evolved through a \ac{CE} episode,
 the \ac{MS} companions are too light to match the observed distribution of Be star masses in \acp{BeXRB} (
 $\gtrsim 6\,\mathrm{M_\odot}$). Even when there is a second episode of stable mass transfer following the re-expansion of the stripped primary after the helium main sequence \citep{Laplace2020}, its short duration is unlikely to allow for significant accretion onto the \ac{MS} star. 
 These findings are consistent 
   with the  conclusions of \citet{presentation}, who similarly considered and discarded stable mass transfer following a common envelope phase as a possible formation channel for Be stars in \acp{BeXRB}. 
 Moreover, systems experiencing a \ac{CE} phase are significantly hardened by it and are therefore characterised by orbital periods that are too short (mostly below $\sim 20$ days) compared with observations. 

This leaves stable mass transfer as the most likely channel for producing \acp{BeXRB}, which is consistent with, e.g., \citet{Pols1991, portegies1995formation, vanBever1997, presentation}.

\begin{figure}
\begin{centering}
\includegraphics[width=8.5cm] {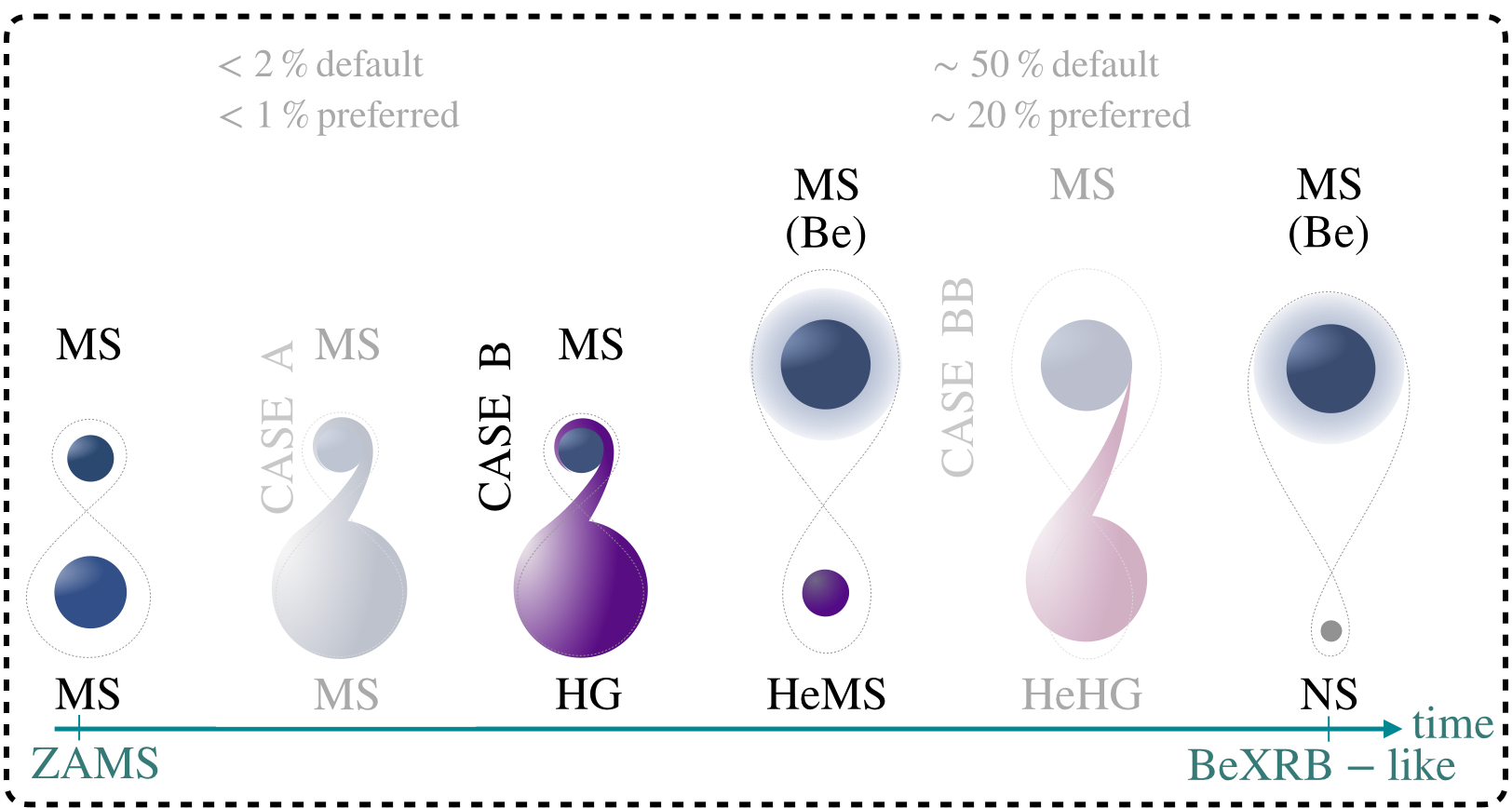}
\caption{Schematic representation of our preferred \ac{BeXRB} formation channel. All the mass transfer episodes are stable. The symbols represent the evolutionary stages of each star in the binary: main-sequence ($\mathrm{MS}$); Hertzsprung gap ($\mathrm{HG}$);  Helium main-sequence ($\mathrm{HeMS}$); Helium Hertzsprung gap ($\mathrm{HeHG}$) and neutron star ($\mathrm{NS}$). After the mass transfer episode from a Hertzsprung-gap primary, the accreting secondary spins up and  becomes a Be star in our model. The reduced-contrast sketches refer to evolutionary phases (Case A and Case BB mass transfer) experienced by only a fraction (percentages above the sketch) of the total \ac{BeXRB}-like population. 
}
\label{fig:FormationChannel}
\end{centering}
\end{figure}

\subsubsection{\ac{BeXRB}-like systems}
\label{sec:BeXBlike}
Motivated by these considerations, we define \ac{BeXRB}-like systems in our simulations as binaries:
\begin{enumerate}
\item composed of a \ac{NS} and a \ac{MS} star with mass $M_{\mathrm{MS}}\gtrsim 3\, \mathrm{M_{\odot}}$. 
This mass cut roughly selects stars of B \& O spectral types \footnote{Late B stars can actually extend to slightly lower masses, $\gtrsim 2\,\mathrm{M_\odot}$ \citep{martayan2007effects,Huang2010}); however, extending the \ac{MS} mass range to such low values is irrelevant for our investigation.} \citep{belczynski2009apparent,grudzinska2015formation};
\item where the \ac{MS} star is not overflowing its Roche lobe,
as this would likely destroy the decretion disk \citep{Panoglou2016};
\item whose secondary accreted during dynamically stable mass transfer from a hydrogen-rich primary (see above).
\end{enumerate}
These conditions do not explicitly require the formation of a Be star in the system.  Our current lack of understanding of the mechanisms which  generate the decretion disk prevents us from identifying such systems.  We therefore limit our definition of \ac{BeXRB}-like systems to the points listed above.  We later discuss the occurrence of Be versus B stars by comparing our simulated population with the observed \ac{BeXRB} sample.

\subsubsection{The history of \ac{BeXRB}-like systems}
In this section we present in more detail the sequence of \acp{SN} and 
mass transfer episodes which lead to the formation of \ac{BeXRB}-like systems. 
Hereafter mentions of mass transfer always refer to Roche lobe overflow from the primary before its collapse into a neutron star.
We will indicate these mass transfer episodes preceding the \ac{BeXRB} stage by the stellar type of the primary at the time of mass transfer, an arrow toward the secondary, and the stellar type of the secondary, e.g., $\mathrm{HG\rightarrow MS}$ for mass transfer from a \ac{HG} primary onto a \ac{MS} secondary.

In figure \ref{fig:FormationChannel} we show our preferred \ac{BeXRB} formation channel, indicating the fraction of the simulated \ac{SMC} population of binaries undergoing multiple mass transfer episodes. 
Similarly to \citet{Pols1991}, we find that all selected binaries 
have experienced mass transfer initiated by the primary after hydrogen exhaustion (case B mass transfer). 
In particular, according to our default model, the primary of \ac{BeXRB}-like systems always overflows its Roche lobe 
at the \ac{HG} stage ($\mathrm{HG\rightarrow MS}$).
More evolved donors are ruled out by our stability criteria, which favour dynamically unstable mass transfer for giants that develop a deep convective envelope \citep{Soberman1997}.

About $\lesssim 10\%$ of our simulated intrinsic population of \acp{BeXRB} start mass transfer while the primary is burning hydrogen in its core (case A mass transfer) at the \ac{MS} stage ($\mathrm{MS\rightarrow MS}$).
However, after re-weighting by the \ac{SMC} \ac{SFR} history to obtain the currently observable population (according to equation \ref{eq:integral}), this fraction becomes negligible (<2\%).

In stellar evolution codes based on the fitting formulae of \citet{hurley2000comprehensive}, a star acquires a core only at the beginning of the \ac{HG}. The mass of this core is estimated from the current total mass of the star and it is therefore lighter for stars which already lost a significant amount of mass during the \ac{MS}. The core mass, in turn, is one of the key parameters which determine the fate of a star.  Therefore, in order to retain a sufficiently massive core to form a \ac{NS} after engaging in $\mathrm{MS\rightarrow MS}$ transfer, the primary must have been quite massive, and likely had a similarly massive companion given the uniform distribution of the initial mass ratio.  The high masses, in turn, lead to short lifetimes; massive systems born during the peak of the \ac{SMC} \ac{SFR} would no longer be detectable as \acp{BeXRB}.  This explains the rarity of such systems in the simulated \ac{SMC} population.

A considerable fraction of binaries in our simulated \ac{BeXRB} populations (in our default model: $\sim 20\%$ of the intrinsic population and $\sim 50\%$ of the \ac{SMC} one) experience a second mass transfer episode after the primary has been stripped by its hydrogen envelope (case BB mass transfer). 
Our analysis may even underestimate the actual incidence of case BB mass transfer.
Recent simulations \citep{Laplace2020} suggest that the radial expansion of partially stripped stars is greater than assumed in codes based on the \citet{hurley2000comprehensive} fitting formulae, like COMPAS.
This case BB mass transfer episode, which in the case of \ac{BeXRB} formation always involves a $\mathrm{MS}$ accretor, happens during the primary's \ac{HeHG} stage, after the primary completes its helium main sequence evolution 
($\mathrm{HeHG\rightarrow MS}$). 
In these cases, we assume that the primary becomes an ultra-stripped star and if it explodes in a \ac{SN}, it experiences lower natal kicks (\ac{US}\ac{SN}).
Together with the typically low orbital separations, these reduced kicks make the disruption of the binary during a \ac{SN} very unlikely, in contrast to the general population of \acp{SN}, leaving behind low-mass remnants \citep{2019Renzo,Eldridge2011}.
Therefore, these systems are prominent in the simulated intrinsic population of \acp{BeXRB}.
Systems experiencing case BB mass transfer are also favoured by re-weighting by the \ac{SMC} \ac{SFR} according to equation \ref{eq:integral}. 
Relatively low initial mass ratios $q_{\mathrm{HG}} = M_{\mathrm{acc}}/M_{\mathrm{donor}}\lesssim 0.4$ allow the systems to tighten during the first mass transfer, making subsequent case BB mass transfer more likely.
This difference between the masses of donors and accretors yields lighter \ac{MS} companion stars (due to both their initially lower masses and generally less efficient accretion). In turn this entails a longer duration of the \ac{BeXRB} phase and larger formation time, which matches the time since the peak of the \ac{SMC} \ac{SFR}.
However, the predicted orbital period and \ac{MS} star mass distributions depend only very weakly on the details of case BB mass transfer. This is mostly because of the short time scales and the relatively small amount of mass loss/exchange.

\subsection{Comparisons with observations}\label{sec:comparisons}
\begin{figure}
\begin{centering}
\includegraphics[width=8.45cm]{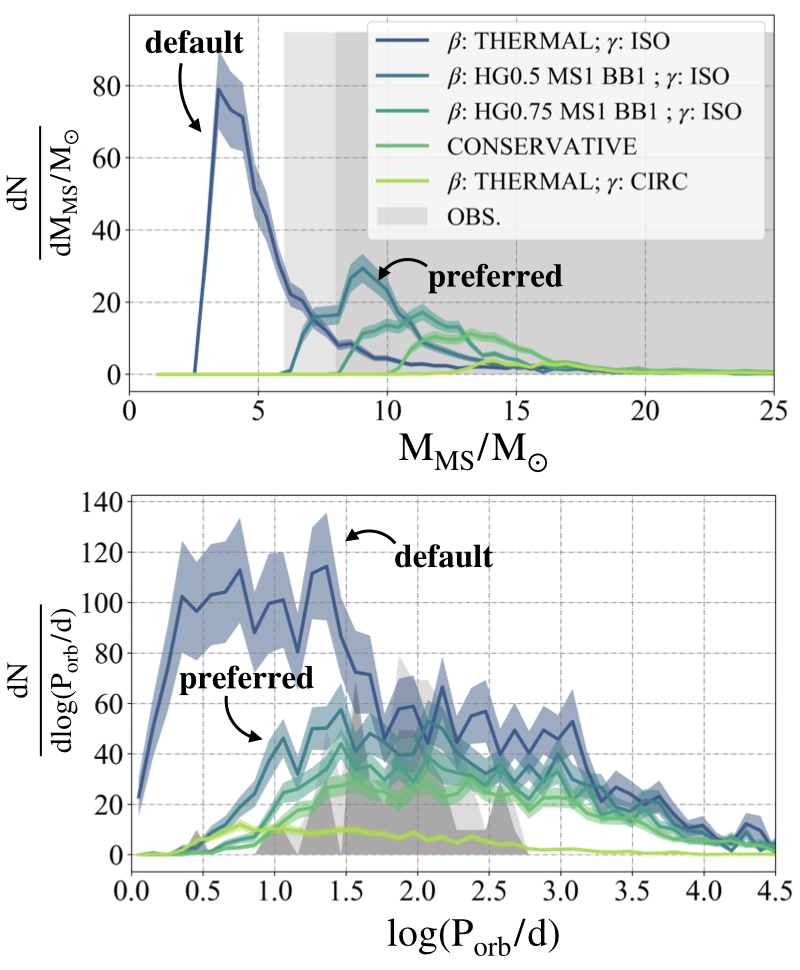}%
\caption{
The impact of mass transfer efficiency: mass (top panel) and orbital period (bottom panel) distributions 
in our simulated \ac{SMC} \ac{BeXRB}-like systems. Different colours correspond to different mass transfer models; shading indicates $1$ $\sigma$ uncertainties.  The grey areas indicate the observed mass range and orbital period distribution of Be stars  in \ac{SMC} \acp{BeXRB}. The dark (light) grey in the top panel corresponds to the common (conservative) threshold of $M_{\mathrm{Be\,in\, BeXRBs}}\geq 8\mathrm{M_\odot}$ ($\geq 6\mathrm{M_\odot}$). The dark (light) grey in the bottom panel corresponds to the 44 (25) systems listed in the \citealt{Coe&Kirk2015} catalog with a measured (inferred from the spin period via equation~\ref{eq:Corbet})  orbital period. 
}
\label{fig:MMS_main}
\end{centering}
\end{figure}
We compare observations to our simulated \ac{SMC} population of \acp{BeXRB} in figures \ref{fig:MMS_main} to \ref{fig:stability} and in table \ref{tab:N}. 
When we plot \ac{MS} mass and orbital period distributions,
we label our variations showing the assumed models for accretion efficiency and specific angular momentum loss. 
We also explicitly highlight our default model (see section \ref{subsection:masstransfer}) and preferred variation. 
Our preferred model assumes fixed accretion efficiencies ($\beta= 0.5$, 
for all mass transfer phases, except for the fully conservative case A and BB mass transfer and the Eddington limited accretion onto a compact object, see section \ref{sec:beta}) 
and isotropic re-emission of non-accreted material from the surface of the accretor, as justified below.  This model is also identified as $\mathrm{\beta:\, HG0.5\, MS1\, BB1;\, \gamma: ISO}$.

\subsubsection{Be star mass distribution}
\label{sec:Be_masses}
The distributions of \ac{MS} star masses in \ac{BeXRB}-like systems are shown in the top panel of figure \ref{fig:MMS_main}; we also show the orbital period distribution associated to each of these simulated \ac{SMC} populations in the bottom panel.
In figure \ref{fig:MMS_main}, 
we show the impact of the accretion efficiency $\beta$, by varying it from the $\mathrm{THERMAL}$ prescription to different fixed values: 0.5, 0.75, and 1 (CONSERVATIVE), as described in \ref{sec:beta} (hereafter shown as $\mathrm{\beta: HG})$. 

For the formation of our \ac{BeXRB}-like systems, the relevant mass transfer phases are: $\mathrm{MS\rightarrow MS}$, $\mathrm{HG\rightarrow MS}$ and $\mathrm{HeHG\rightarrow MS}$, with $\mathrm{HG\rightarrow MS}$ as the most significant mass transfer phase. 
Because our results only very weakly depend on $\mathrm{MS\rightarrow MS}$ and $\mathrm{HeHG \rightarrow MS}$ mass transfer episodes, they are always assumed to be conservative when a fixed $\beta$ value is applied.
 This is shown by the labels $ \mathrm{\beta:\, MS1}$ and  $\mathrm{\beta:\, BB1}$. 
These assumptions allow us to be conservative in our inference over the accretion efficiency during the $\mathrm{HG\rightarrow MS}$ mass transfer when comparing the predicted Be star mass distributions with the observed ranges. 

All  populations of simulated \ac{SMC} \ac{BeXRB}-like in figure \ref{fig:MMS_main} assume that, during the mass transfer episodes preceding the \ac{BeXRB}-like phase, any material lost from the binary was ejected from the surface of the accretor ($\mathrm{\gamma: ISO}$). 
This isotropic re-emission model better predicts the orbital distribution of \acp{BeXRB} (see below). 
However we also report a single case where $\mathrm{\beta: THERMAL}$ and $\mathrm{\gamma: CIRC}$, to show how increasing the angular momentum loss leads to heavier \ac{MS} stars, 
independently of the assumed accretion efficiencies.

The blue and dark green lines in figure \ref{fig:MMS_main} show our default and preferred models. Dark and light grey in the top panel mark the common and conservative ranges for Be star masses in \acp{BeXRB} $M_{\mathrm{Be\, in\, BeXRBs}}\geq 8\mathrm{\, M_\odot}$  and $\geq 6\mathrm{\, M_\odot}$, respectively.  Dark and light grey in the bottom panel mark the directly measured orbit periods of observed \acp{BeXRB} in the CK catalog and those inferred from spin periods by assuming the Corbet relation, equation \ref{eq:Corbet} in section \ref{sec:SMC}, respectively.

According to our default \ac{SMC} model, most of the detected \acp{BeXRB} should have a \ac{MS} companion $M_{\mathrm{MS}}< 6\, \mathrm{M_\odot}$. This low peak mass clearly contradicts observations.   
Because synthetic \acp{BeXRB} with low-mass companions do not cluster toward high orbital separations, selection effects are unlikely to resolve this incompatibility. 

\begin{figure}
\begin{centering}
\includegraphics[width=8.75cm] {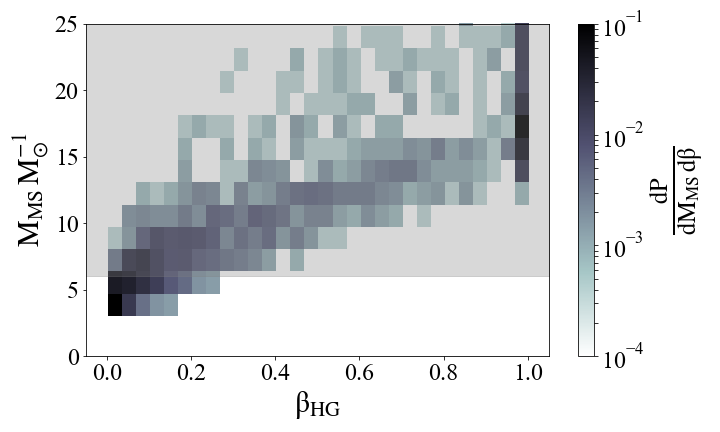}
\caption{
Simulated main-sequence companion (Be star) mass $M\mathrm{_{\mathrm{MS}}}$ versus the efficiency of mass transfer for a Hertzsprung-gap donor $\mathrm{\beta_{\mathrm{HG}}}$.
The colour bar shows the intrinsic probability, for \ac{BeXRB}-like systems, of obtaining a specific combination of $M\mathrm{_{MS}}$ and $\mathrm{\beta}$ according to our default model.
As in figure \ref{fig:MMS_main}, the grey area shows the range of observed Be star masses, according to our conservative interpretation \citep{Hohle2010}.
}
\label{fig:effMTvsMMS}
\end{centering}
\end{figure}

Figure \ref{fig:effMTvsMMS} displays the joint probability distribution of \ac{MS} companion masses and mass transfer efficiencies of Hertzsprung-gap donors for \ac{BeXRB}-like systems formed in our default model.  This clearly shows that the main determining factor for the companion mass distribution is the accretion efficiency $\beta$ during the $\mathrm{HG\rightarrow MS}$ mass transfer.
 
For the simulated intrinsic \ac{BeXRB}-like systems, our THERMAL assumption generates two peaks in the distribution of the $\mathrm{HG\rightarrow MS}$ accretion efficiency $\beta_\mathrm{HG}$: a narrow peak at $\sim 1$, and a broad peak concentrated at values $\beta_\mathrm{HG}\lesssim 0.2$. 
This second peak
is the cause of the inconsistency between 
the predictions of our default model and the observed Be star masses in \acp{BeXRB}. 
Lower $\beta_{\mathrm{HG}}$ values reduce the accretion of the primary hydrogen envelope onto the secondary, leaving the \ac{MS} mass of initially light stars almost unchanged. 
Binaries containing low-mass \ac{MS} stars then dominate our simulated \ac{SMC} population (see top panel of figure \ref{fig:MMS_main}) because they are favoured by both the \ac{IMF} and the re-weighting for the lifetime of the \ac{BeXRB} stage. 

As shown in the top panel of figure \ref{fig:MMS_main}, applying progressively higher accretion efficiencies during the $\mathrm{HG\rightarrow MS}$ mass transfer gradually moves the \ac{MS} star mass distribution toward higher masses that better match the observed range. 
In particular, imposing half-conservative $\mathrm{HG\rightarrow MS}$ mass transfer already shifts the entire mass distribution above $6\,\mathrm{M_\odot}$. 
This result is consistent with figure \ref{fig:effMTvsMMS}, which suggests that 
to match observations, 
the overwhelming majority of \acp{BeXRB} must have experienced a post-\ac{MS}$\rightarrow$\ac{MS} mass transfer with accretion efficiency above $\gtrsim 0.3$. 

The $\mathrm{\beta:\, THERMAL;\, \gamma:\, CIRC}$ model (lightest-green line in the top panel of figure \ref{fig:MMS_main}) 
confirms the findings of \citet{portegies1995formation}, who proposed high angular momentum loss to explain the Be mass distribution in \acp{BeXRB}.
\citet{portegies1995formation} suggested matter leaving from the second Lagrangian point or, similarly, with a specific angular momentum 6 times higher than the binary's one ($\gamma = 6$). 
In our simulations, this last case qualitatively resembles the results obtained for the circumbinary ring mass-loss mode, which yields $\gamma=6$ for $q \sim 0.6$. 
In these models, the high angular momentum loss disfavours low mass ratio systems, which are now very likely to merge. 
Consequently, the masses of the \ac{MS} companions 
in the surviving binaries of our high angular momentum loss model ($\mathrm{\beta:\, THERMAL;\, \gamma:\, CIRC}$) match well the observed mass range of Be stars in \acp{BeXRB}.

\subsubsection{Orbital period distribution}
\label{sec:Porb}

\begin{figure}
\begin{centering}
\includegraphics[width=8.45cm] {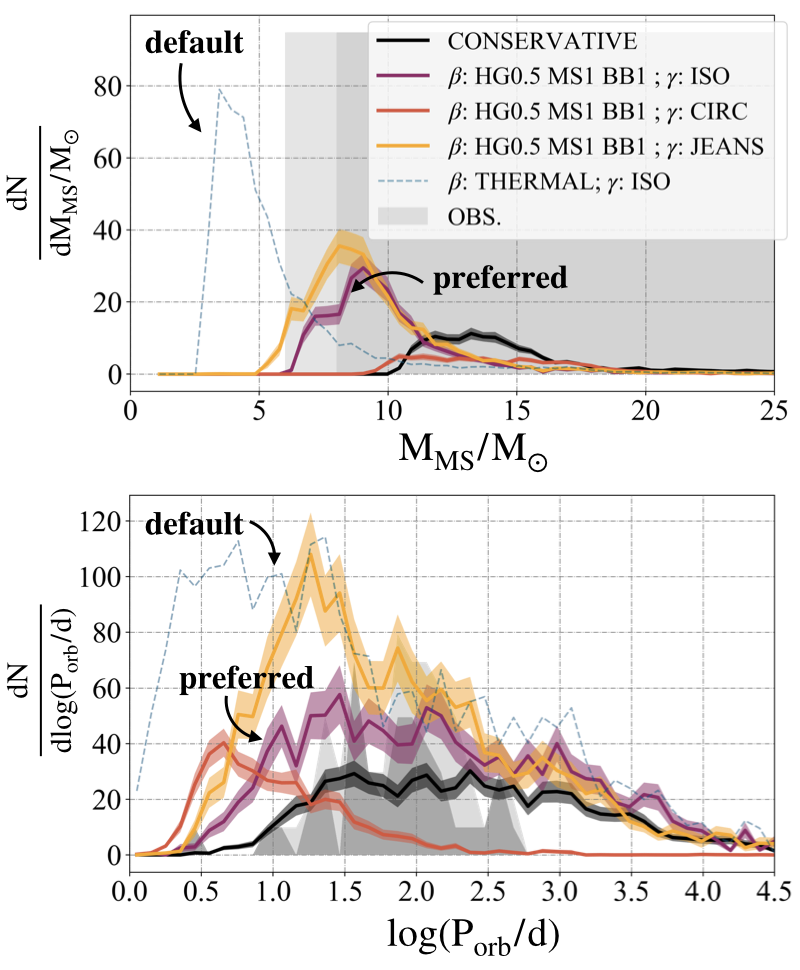}%
\caption{
The impact of angular momentum loss: mass (top panel) and orbital period (bottom panel) distributions 
in our simulated \ac{SMC} \ac{BeXRB}-like systems. Different colours correspond to different mass transfer models; shading indicates $1$ $\sigma$ uncertainties.  The grey areas indicate the observed mass range and orbital period distribution of Be stars  in \ac{SMC} \acp{BeXRB}. The dark (light) grey in the top panel corresponds to the common (conservative) threshold of $M_{\mathrm{Be\,in\, BeXRBs}}\geq 8\mathrm{M_\odot}$ ($\geq 6\mathrm{M_\odot}$). The dark (light) grey in the bottom panel corresponds to the 44 (25) systems listed in the \citealt{Coe&Kirk2015} catalog with a measured (inferred from the spin period via equation~\ref{eq:Corbet}) orbital period. 
}
\label{fig:Porb_main_MMS_secondary}
\end{centering}
\end{figure}

In the previous subsection, we showed that the \ac{MS} mass distribution in \ac{BeXRB}-like systems is primarily determined by the mass transfer efficiency.  Here, we analyse the orbital period distribution of these systems, and demonstrate its sensitivity to the angular momentum carried away during non-conservative mass transfer.

Figure \ref{fig:Porb_main_MMS_secondary} shows the dependence of the Be star masses (top panel) and binary orbital periods (bottom panel) of \ac{BeXRB}-like systems in simulated \ac{SMC} populations.  We assume constant accretion efficiencies $\mathrm{\beta:\, HG0.5\, MS1\, BB1}$ and explore different angular momentum models. 
With this assumption for the accretion efficiencies, the masses of the \ac{MS} companion at the \ac{BeXRB} stage are consistent with the observed range and depend only slightly on the 
 angular momentum loss models.
For comparison, we also plot the orbital period distribution for fully conservative mass transfer (black line) and for our default model (dashed blue line). 
In dark grey we show the distribution of the 44 observed orbital periods reported in the CK catalogue. In light grey we show the orbital period distribution obtained by applying the Corbet relation fit from equation (\ref{eq:Corbet}) to the 25 systems in the CK catalogue which only have measured spin periods.  The width of the shaded regions around the curves again indicates combined 1-$\sigma$ uncertainties from the \ac{SMC} \ac{SFR} history and COMPAS sampling; the sizes of the spikes in the plots are consistent with these statistical fluctuations.

The orbital period distribution from our default model (dashed blue line) considerably differs from observations. 
The predicted overabundance at long orbital periods can be explained by observational selection effects (see section \ref{sec:selection_effects}). 
However, the predicted but unobserved 
peak of \acp{BeXRB} at short orbital periods indicates a failure of this model.
This peak is mostly due to the population of systems experiencing \ac{US}\acp{SN} (see e.g. figure \ref{fig:Temperature_US_SS}).
As mentioned above, these binaries are very likely to survive the \ac{SN} and are characterised by small separations and low masses.
In our default model, this population is only partially suppressed by the stability criteria \citep{Podsiadlowski1992}, which allow all systems with $q_{\mathrm{HG}}\gtrsim 0.23$ to engage in stable $\mathrm{HG\rightarrow MS}$ mass transfer.

In purple we show the orbital period distribution under the assumption of fixed mass transfer efficiency $\mathrm{\beta:\, HG0.5\, MS1\, BB1}$ and isotropic re-emission from the surface of the accretor (our preferred model). 
Both the preferred and fully conservative mass transfer (black) models display orbital period distributions which range and peak at values similar to the observed ones. 
Neglecting for now the total number of predicted \ac{BeXRB}-like systems, represented by the area below the curve, both these models predict orbital period and \ac{MS} mass distributions in general agreement with observations. 

The orange curves in figure \ref{fig:Porb_main_MMS_secondary} describe the population of simulated \ac{SMC} \ac{BeXRB}-like systems which experienced mass transfer with matter leaving the binary from a circumbinary ring, $\mathrm{\gamma:\, CIRC}$ (see equation \ref{eq:gamma_circ}). 
The expected orbital period distribution in this model peaks at a few days, in disagreement with observations. 
This shift toward 
shorter separations relative to our preferred model is due to the greater loss of angular momentum during mass transfer. 
High angular momentum losses are therefore unlikely to be responsible for massive \ac{MS} stars in \acp{BeXRB}.  
These considerations also strongly disfavour more extreme angular momentum losses, as would be the case for matter leaving the binary from L2 or L3. 
Similar conclusions can also be inferred from the circumbinary ring mode combined with the THERMAL prescription as shown in the bottom panel of figure \ref{fig:MMS_main}.

In yellow, we plot the orbital period distribution for the \ac{BeXRB} population obtained assuming the Jeans mode of mass loss.  During the first mass transfer episode $\gamma_\mathrm{JEANS}$ is always less than one, meaning the binary increases its specific angular momentum by losing matter and so it typically widens.
Contrary to isotropic re-emission from the accretor surface, the more unequal the component masses, the less angular momentum is assumed to leave the system. 
This scenario is therefore appealing for moving the short period systems found in our default model toward higher separations.
However, according to our stability criteria, this combination of $\beta$ and $\gamma$ values allows binaries with mass ratio as low as $0.14$ (at the onset of mass transfer) to engage in stable $\mathrm{HG\rightarrow MS}$ mass transfer. 
Most of these low $q$ systems end up contributing to the short orbital period tail after their primaries experience an \ac{US}\ac{SN}, leading to an orbital period peak around $\sim 20$ days, too low to match the observations.

Alternatively, matter lost during mass transfer episodes could leave the binary with the specific angular momentum of the system. This assumption corresponds to fixing $\gamma$ to 1 and is applied in other population synthesis codes, such as StarTrack (e.g. \citealt{belczynski2009apparent,dominik2013double}). 
In combination with the accretion efficiency 
$\mathrm{\beta:\, HG0.5,\, MS1,\, BB1}$ (or similarly $\beta=0.5$ for all mass transfer), this angular momentum loss model produces similar results to Jeans mode mass loss, with orbital periods peaking at $\sim [10-30]\,$d, too short compared to observations. 

For both Jeans mode and $\gamma=1$ mass loss, applying reduced \ac{US}\ac{SN} kicks only to stars whose helium envelope was stripped by a \ac{NS} \citep{vigna2018formation} rather than the broader range of stars discussed in section \ref{sec:SN}, mitigates the inconsistencies between the predicted orbital period distribution and observations.   We discuss these findings in the context of \ac{US}\acp{SN} and case BB mass transfer in section \ref{sec:disc_shortP}. 

Many of our simulated \ac{SMC} populations of \ac{BeXRB}-like systems over-predict the number of binaries with orbital period longer than $\sim 500\,$ days. 
This tail is currently difficult to test observationally, as observational selection effects work strongly against the detection of these systems (see section \ref{sec:selection_effects}). 
Systems at large separation may however be partially probed through radio pulsar observations. 
Our models predict 20--40 \ac{BeXRB}-like systems in the \ac{SMC} at orbital periods longer than the maximum measured of $\sim520$~days. At these large separations, we expect \acp{NS} to rarely (at periastron passage for highly eccentric orbits) or never accrete from the decretion disk of the companion, and so to possibly remain radio active for most of the typical pulsar lifetime of $\sim 10^7$~yrs \citep[e.g.][]{Lorimer2008}. According to our models, these long orbital period \acp{BeXRB} typically survive as \ac{NS}+\ac{MS} star systems for $\sim 7\times 10^6$~yrs. This suggests that all our predicted long period \ac{BeXRB}-like systems may currently host radio pulsars, which could be detected by instruments such as the Square Kilometer Array \citep[e.g.][]{Weltman2020}.
The number of foreseen detections should nonetheless account for the beaming angle, which may reduce the observable pulsars to a few (only $\sim$20\% of pulsars are detectable according to \citealt{Lorimer2008}), and possible intermittence of the radio activity for systems in eccentric orbits. 
 Indeed a subpopulation of these systems may be intermittent radio pulsars, similarly to PSR B1259-63, which has a 3.4 year orbital period and shows gamma-ray flares at periastron passages (e.g. \citealt{Miller-Jones2018}). 
  
From the point of view of our binary evolution models, 
the long orbital period systems could be suppressed by increasing the overall distribution of natal kick velocities assigned to \acp{NS} born in \ac{EC}\acp{SN} (see figure \ref{fig:Temperature_US_SS}). 
This approach, however, would also considerably affect other predictions. Of particular relevance in this context are predictions concerning \acp{BNS}, whose progenitors often experience a \ac{BeXRB}-like phase with relatively long orbital periods \citep{vigna2018formation}.

The imperfectly known relationship between the carbon-oxygen core mass and the post-supernova remnant mass \citep[and natal kick, see, e.g.,][]{MandelMueller:2020} will affect the orbital period distribution inferred in our study.  However, the relatively limited range of \ac{NS} masses means that the impact on individual orbital periods should not exceed a $\sim \sqrt{2}$ level.  This is an active topic of ongoing research and is further discussed in \citet{vigna2018formation,Mandel:2020} in the context of COMPAS.

\subsubsection{Estimated number of simulated \ac{SMC} \ac{BeXRB}-like systems}
In table \ref{tab:N} we report the number of \ac{SMC} \ac{BeXRB}-like systems in our simulations with orbital period below $\sim 520$ days (the longest orbital period in the CK catalogue), as a function of the different accretion efficiency $\beta$ (columns) and angular momentum loss $\gamma$ (rows) models.
There are three main effects that influence the predicted number of \ac{BeXRB}-like systems as a function of $\beta$ and $\gamma$: (i) the growth in accretor masses, which impacts the duration of the \ac{BeXRB}-like phase, (ii) the difference in orbital evolution and (iii) the stability of $\mathrm{HG\rightarrow MS}$ mass transfer (see table \ref{tab:qcrit}).

The first row in table \ref{tab:N} shows the 
expected number of \acp{BeXRB}, assuming isotropic re-emission from the surface of the accretor ($\mathrm{\gamma:\, ISO}$). 
Increasing the typical accretion efficiency from the THERMAL prescription to the fully conservative case, we notice a gradual drop in the expected number of simulated \ac{SMC} \ac{BeXRB}-like systems. 
This is also shown in figure \ref{fig:MMS_main}, where the area below the curve, representing the total number of \ac{BeXRB}-like systems in our simulated \ac{SMC} population, shrinks for progressively more conservative mass transfer.
There are several reasons which explain this trend. 
First, the growth in the mass of the \ac{MS} secondary star through accretion reduces the duration of the \ac{BeXRB} stage.
Secondly, more conservative mass transfer leads to larger typical separations, so the binaries are more likely to be disrupted by the \ac{SN} kicks.
Larger separations after the $\mathrm{HG\rightarrow MS}$ mass transfer also imply that fewer binaries engage in case BB mass transfer.
In turn, this leads to fewer primaries experiencing the reduced \ac{US}\ac{SN} kicks and so fewer systems surviving the \ac{SN} explosion. 
Finally, assuming the isotropic re-emission model for the specific angular momentum loss, the response of the Roche lobe to mass transfer favours more unstable interactions for higher $\beta$ values. 
For these reasons, our preferred model predicts fewer \acp{BeXRB} than the default and shows a drastic decrease in the number of systems with orbital periods below $\sim 30$ days. 

As mentioned in section \ref{sec:Be_masses}, high angular momentum losses could effectively explain the high masses of Be stars in \acp{BeXRB}  \citep{portegies1995formation}. 
However, as shown in figures \ref{fig:MMS_main}, \ref{fig:Porb_main_MMS_secondary} and in the second row of table \ref{tab:N}, the expected number of \ac{BeXRB}-like systems drops considerably below the 69 observed \acp{BeXRB} listed in the CK catalogue.
This is both due to the increased number of mergers and our stability criteria, which in the circumbinary ring mode disfavour stable $\mathrm{HG\rightarrow MS}$ mass transfer, especially if the mass transfer is highly non-conservative, as is often the case for our default $\mathrm{THERMAL}$ $\beta$ model (see table \ref{tab:qcrit}). 
Together with the orbital period distributions, the low numbers of predicted \ac{BeXRB}-like systems therefore rule out the circumbinary ring mode (if located at twice the binary's separation) for $\mathrm{HG\rightarrow MS}$ mass transfer. 

The third row of table \ref{tab:N} shows the expected number of \ac{BeXRB}-like systems assuming that the matter lost during mass transfer leaves the system from the surface of the donor. 
The high predicted numbers mostly stem from the stability criteria applied for $\mathrm{HG \rightarrow MS}$ mass transfer. Table \ref{tab:qcrit} shows that for accretion efficiencies $\lesssim 0.5$, the Jeans mode assumption allows systems with very unequal masses ($q_\mathrm{crit}\lesssim 0.14$) to experience stable mass transfer. This increases the number of binaries which experience the reduced kicks of \ac{US}\acp{SN}, and therefore populates the binaries at short orbital periods. 
As mentioned in section \ref{sec:formationCH}, binaries which experience these reduced kicks are also often characterised by lower \ac{MS} star masses
(this is also visible for $\mathrm{\beta:\,HG0.5,\, MS1,\, BB1}$ in the top panel of figure \ref{fig:Porb_main_MMS_secondary}, by comparing the distributions obtained for Jeans mode vs.~isotropic remission mass transfer). 
Lower masses for the \ac{MS} companions then affect the overall number of predicted \ac{BeXRB}-like systems, by extending the lifetime of the \ac{BeXRB}-like phase. 
If we assume the THERMAL prescription for the accretion efficiency, the contribution of binaries experiencing \ac{US}\acp{SN} is, however, less prominent than for the case of isotropic re-emission. 
This is due to the orbital evolution, which tends to bring low mass ratio binaries closer together for $\mathrm{\gamma:\, ISO}$. 
Moreover, within the THERMAL prescription, the advantage of the Jeans mode coming from the lower critical mass ratio is almost negligible. 
For example, if $q=0.15$ and the secondary has a mass above $3\mathrm{M_\odot}$, the primary donor would be more massive than $20\mathrm{M_\odot}$, i.e., may be too heavy to produce an \acp{NS} and hence a \ac{BeXRB}-like systems. 

\begin{table}
\centering
\begin{tabular}{|l|c|c|c|}\hline
\diaghead{\theadfont JEANS}%
  {$\gamma$}{$\beta$}&\thead{THERMAL}&\thead{HG0.5 MS1 BB1}& \begin{tabular}{@{}c@{}}CONSER- \\ VATIVE\end{tabular}\\
\hline
{ISO} & $ 200\pm 20 $ & $ 85\pm 10 $& $45 \pm 5$\\
\hline
{CIRC}  & $ 19\pm 2$ &$  40\pm 5 $ & $45 \pm 5$\\
\hline
{JEANS} & $ 130\pm 15 $ & $135 \pm 15$ & $45 \pm 5$\\
\hline 
\end{tabular}
\caption{The table reports the number of \acp{BeXRB} with orbital period under 520 days (longest period listed in the CK catalogue), predicted in the \ac{SMC} according to our synthetic populations of binaries.
According to the CK catalogue there are at least 69 \acp{BeXRB} in the \ac{SMC}, of which at the very least 44 (all systems with observed orbital periods) have period below 520 days. 
The rows correspond to different modes of angular momentum loss, while the columns represent different accretion efficiency models.}
\label{tab:N}
\end{table}

\subsubsection{Stability criteria: impact on synthetic \ac{BeXRB} populations}
\label{subsec:stability}
\begin{figure}
\begin{centering}
\includegraphics[width=8.75cm] {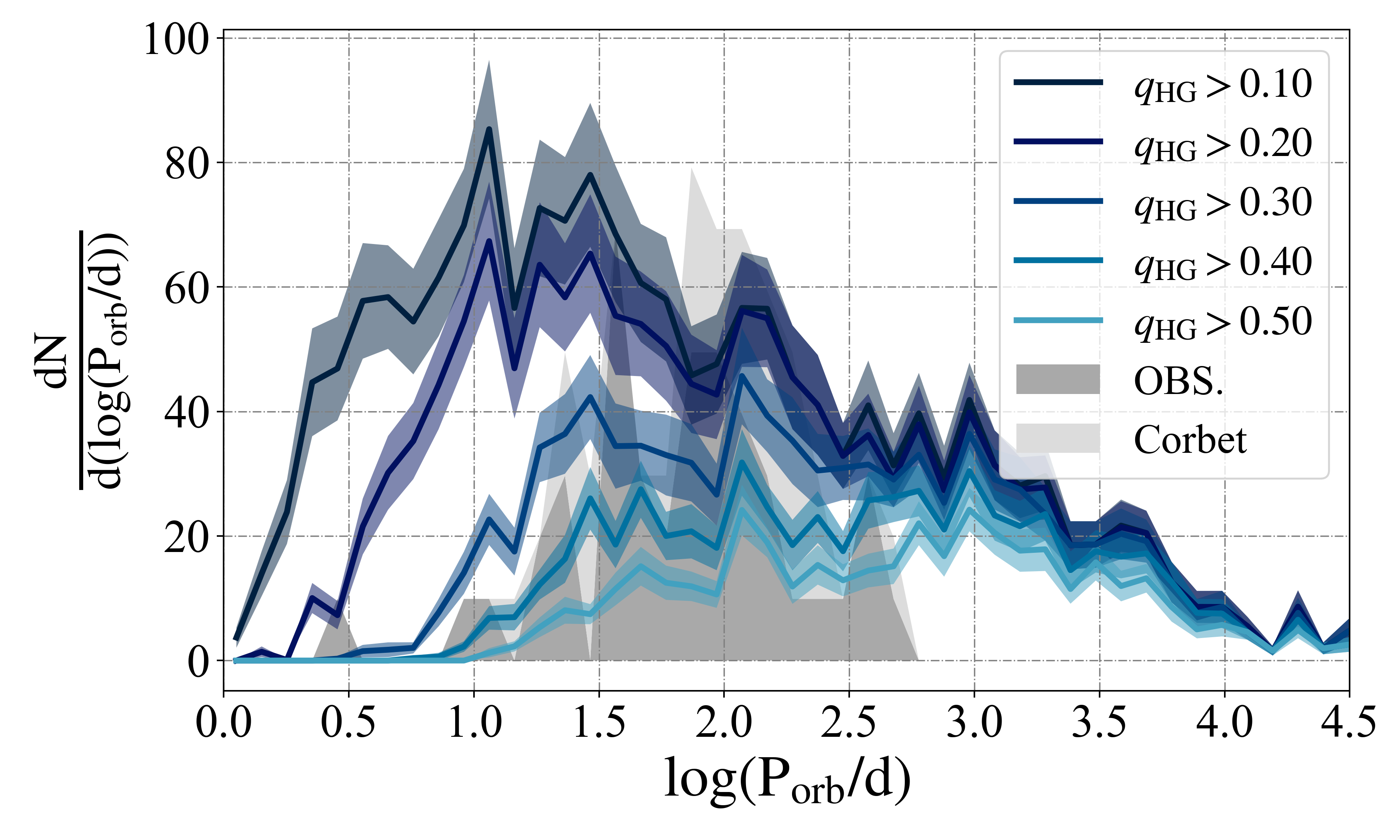}
\caption{The effect of different critical mass ratios (in different shades of blue) on the predicted \ac{SMC} \ac{BeXRB} orbital period distribution and total number of \acp{BeXRB}. We focus on mass transfer episodes from \ac{HG} primaries, as this is the crucial episode in our preferred \ac{BeXRB} formation channel. We show results for the preferred model: constant accretion efficiencies ($\mathrm{\beta: HG0.5\, MS1\, BB1}$) and isotropic re-emission from the surface of the accretor ($\gamma: ISO$).
Under these mass transfer assumptions, our stability criterion for Hertzsprung gap stars ($\zeta^*_{HG} = 6.5$) would normally correspond to a mass ratio of $q_{\mathrm{HG}}\gtrsim0.24$. 
The meaning of shaded areas is the same as in figure \ref{fig:Porb_main_MMS_secondary}.}
\label{fig:stability}
\end{centering}
\end{figure}

The stability criteria adopted for the post-$\mathrm{MS\rightarrow MS}$ mass transfer play a crucial role in the formation of our \ac{BeXRB}-like systems. 
With our default settings only \ac{HG} donors contribute to the population of \ac{BeXRB}-like systems, and assuming isotropic re-emission, this implies a critical mass ratio between $\sim 0.22$ and $\sim 0.26$ (see table \ref{tab:qcrit}).
The stability of mass transfer is, however, highly uncertain. 
In particular, when rapid accretion causes the accretor to swell and over-fill its Roche lobe, producing a co-rotating envelope, the orbit of the binary may shrink due to angular momentum loss from the system
without leading to a classical \ac{CE} spiral-in. Instead, stable mass transfer may commence (see figures 3--5 in \citealt{Podsiadlowski1992}). 
This idea was originally applied to stars with convective envelopes. 
However \citet{Pols1994} and \citet{Wellstein2001} have 
shown that a similar behaviour 
is also expected for most massive \ac{HG} donors if the mass
ratio is less than $\sim 0.7$ (see e.g. figure 5
in \citealt{Pols1994}). 
These systems are all expected to form a contact phase with at least a temporary \ac{CE} in which they lose a lot of angular momentum and may therefore merge. 
\ac{HG} stars may nonetheless be able to avoid a runaway plunge-in  \citep{Pavlovskii2017}. 
The details of this process are not well understood and it is therefore possible that some of the systems that we assume can successfully reach a \ac{BeXRB} stage, should instead have merged. 
Figure \ref{fig:stability} shows the systems that would survive this mass transfer phase under more stringent conditions on the mass ratio.  It indicates that, in order to match the observed number of \ac{SMC} \acp{BeXRB}, only a small fraction of binaries can merge prematurely through this channel.

The critical mass ratio not only determines how many simulated systems can become \ac{BeXRB}-like, but also their distribution in \ac{MS} masses and orbital periods (e.g. \citealt{presentation,Zwart1996}). 
This is shown in figure \ref{fig:stability}, 
which also indicates that the typical critical mass ratio for $\mathrm{HG\rightarrow MS}$ stability should range between $0.2$ and $0.3$ for our treatment of \ac{US}\acp{SN}.

\begin{figure}
\begin{centering}
\includegraphics[width=8.75cm] {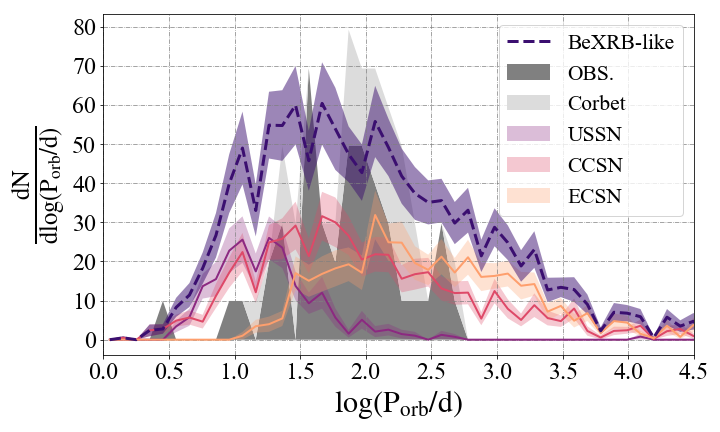}
\caption{Orbital period distribution of \ac{BeXRB}-like systems assuming the alternative stability criterion based on temperature, described in section \ref{sec:stability-method}.
Shaded grey areas follow the description of figure \ref{fig:MMS_main}. Our synthetic \ac{BeXRB}-like systems are divided into three sub-populations, based on the supernova experienced by the \ac{NS} progenitor: the magenta curve tags \ac{US}\acp{SN}, the pink one \ac{CC}\acp{SN} and the orange one \ac{EC}\acp{SN}.}
\label{fig:Temperature_US_SS}
\end{centering}
\end{figure}

In figure \ref{fig:Temperature_US_SS} we show the orbital period distribution of \ac{BeXRB}-like systems obtained by assuming a temperature based stability criterion (see section \ref{sec:stability-method}). 
This variation for determining the stellar response allows early core helium burning stars to engage in stable mass transfer onto a \ac{MS} star and so to contribute to our synthetic population of \acp{BeXRB}.
The simulated \ac{SMC} population of \ac{BeXRB}-like systems in figure \ref{fig:Temperature_US_SS} has been evolved by applying our preferred model of specific angular momentum loss and accretion efficiency ($\mathrm{\beta:\, HG0.5\, MS1\, BB1;\, \gamma:\, ISO}$), under which stable mass transfer from core helium burning donors also has constant $\beta=0.5$. 
We find the results obtained under this assumption to be in good agreement with observations.

\section{Discussion}
\label{sec:discussion}
In the previous sections, we compared the observed \acp{BeXRB} listed in the \ac{SMC} CK catalog against populations of \ac{BeXRB}-like systems evolved by the 
rapid population synthesis code, COMPAS. We identify \ac{BeXRB}-like systems only based on the evolutionary phase of their components (\ac{NS}+\ac{MS} stars) and their binary interaction history (see section \ref{sec:BeXBlike}). 
We do not evolve the rotational velocity of the binary components and we therefore do not define Be stars in terms of the \ac{MS} stars' spin.
We constrain the formation channel of \ac{BeXRB}-like systems and learn about binary evolution by comparing observations with predictions, based on different treatments of mass transfer. Here, we discuss our most relevant findings. 
 
\subsection{Accretion efficiency during stable mass transfer}
The comparison between observations and our synthetic population of \ac{BeXRB}-like systems suggests that our default model underestimates the accretion efficiency during $\mathrm{HG\rightarrow MS}$ mass transfer.
Meanwhile, the simplified fully conservative and half conservative variations allow the simulated Be-star mass distribution to match the observed range.
According to our study, fully conservative mass transfer is however disfavoured by the low predicted number of synthetic \ac{SMC} \ac{BeXRB}-like systems (see also \citealt{presentation}).
Higher angular momentum losses can also explain the Be mass distribution, but they are unable to reproduce the orbital period distribution and the high number of \acp{BeXRB} observed in the \ac{SMC}.

Our findings on accretion efficiency during $\mathrm{HG\rightarrow MS}$ mass transfer are supported by other studies of massive X-ray binaries  \citep[e.g.,][]{presentation,Kaper1995}.
On the other hand, Wolf-Rayet stars in binaries with a \ac{MS} companion strongly suggest highly non-conservative mass transfer \citep{Shao2016, Petrovic2005}.  Interestingly, \citet{deMink2007} point out a weak correlation between orbital period and the efficiency of mass accretion, possibly due to tides and rotationally limited accretion. 
However, as pointed out by \citet{presentation}, rotationally limited accretion entails highly non-conservative mass transfer, which would leave 
most of the \ac{MS} star masses in \ac{BeXRB}-like systems below $\mathrm{6\,M_\odot}$, in contradiction with observations (see figure \ref{fig:effMTvsMMS}). 
This conclusion highlights the importance of studies based on rapid population synthesis, where key but still uncertain parameters, such as the accretion efficiency during mass transfer, can be easily varied and tested.
Indeed, although the increase in available computing power is gradually facilitating detailed evolution of binary grids \citep[e.g][]{Langer2019},
studies based on fitting formula
still make it possible to explore a wider set of possible physical assumptions.

In our default model, very low accretion efficiencies occur if, at the beginning of the mass transfer episode, the donor mass loss timescale and the timescale of the thermal response of the accretor differ by almost two orders of magnitude (as shown by equation \ref{eq:betaTH}), i.e., if their evolutionary stage is very different.  
The need to reduce the rate of occurrence of these low accretion-efficiency events in order to match observations could indicate that such mass transfer events lead to dynamical instability when the accretor overflows its Roche lobe \citep[e.g.,][]{Nariai:1976, Dewi:2006, ivanova2013common}.  However, a more conservative stable mass transfer than predicted by the simplistic default model is likely required to reproduce both the observed number of systems and the Be-star mass distribution.

\subsection{Metallicty or age as the driver of the \ac{SMC} \ac{BeXRB} excess?}
\begin{figure}
\begin{centering}
\includegraphics[width=8.75cm] {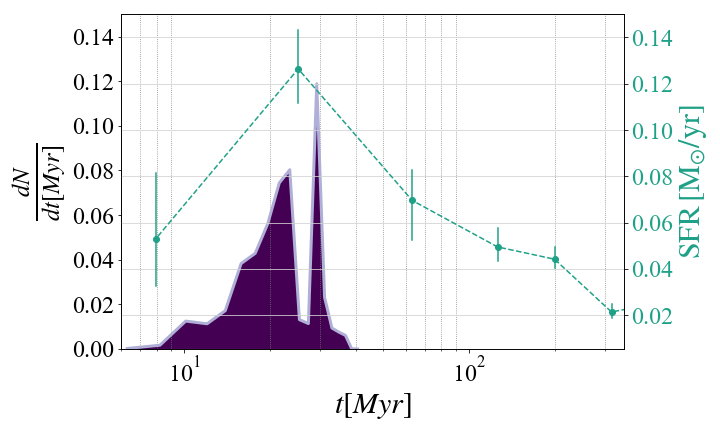}
\caption{In teal, we plot the \ac{SMC} star formation rate as a function of time in Myr, as reported in Rubele et al. 2015 (y-axis on the right). In violet we show the formation time distribution for \ac{BeXRB}-like systems in our default simulation, weighted by the duration of the \ac{BeXRB}-like phase and normalised to 1. 
The formation time distribution peaks around the time of the recent SMC starburst.}
\label{fig:SFR}
\end{centering}
\end{figure}

According to our preferred formation channel, the recent peak of the \ac{SMC} \ac{SFR} ($\sim$20--40 Myr ago) matches the typical time necessary to form \ac{BeXRB}-like systems in our simulations, as shown in figure \ref{fig:SFR}. Similar formation times have been observationally confirmed for \acp{HMXRB} \citep{Williams2013,Williams2018}.
This, together with the strong observational selection effects which impact X-ray detections in the Milky Way, might explain why the number of detected \acp{BeXRB} is so similar in the \ac{SMC} and the Galaxy, despite the very different masses and current \acp{SFR} (see also \citealt{Antoniou2019,Antoniou2010}).

In the general framework of the Be phenomenon, metallicty has also been proven to be a crucial parameter \citep{martayan2007effects}.  In the case of \acp{BeXRB}, our results suggest that all stable $\mathrm{HG\rightarrow MS}$ mass transfer episodes, at and below the \ac{SMC} metallicity, may generate a Be star, which remains rapidly rotating until the end of its \ac{MS} stage.
Assuming this is true, i.e. assuming for simplicity that all \ac{MS} stars in \ac{BeXRB}-like systems are Be stars, the effect of metallicity reduces to its impact on binary evolution.

In terms of the formation of \ac{BeXRB}-like systems, the major impact of metallicity is on stellar  expansion during the \ac{HG} and \ac{HeHG} phases. Greater expansion during the \ac{HG} phase allows $\mathrm{HG\rightarrow MS}$ mass transfer for systems whose components are further apart and so increases the population of \ac{BeXRB}-like systems with longer orbital periods.  Meanwhile, greater expansion of post-helium main sequence stars that were stripped earlier in their evolution increases the population of \ac{BeXRB}-like systems experiencing and surviving \ac{US}\acp{SN}. 

To test the sensitivity of our synthetic \ac{BeXRB} population to metallicity, we repeated the same analyses assuming either half the \ac{SMC} metallicity or the Galactic metallicity.  The overall change in the predicted number of \ac{BeXRB}-like systems is $\lesssim 15\%$ assuming fully conservative mass transfer and $\lesssim 25\%$ for our default model.  According to our default model 
higher metallicity results in a larger population of synthetic \acp{BeXRB}.  Conversely, the highest numbers of \acp{BeXRB}-like systems are reached at the lowest metallicities if assuming fully conservative mass transfer. 

Finally metallicity may influence the lifetime of the \ac{BeXRB}-like systems through wind and decretion disk mass loss of the \ac{MS} star. 
Our results however strongly suggest that most of the \ac{MS} companions in \ac{BeXRB}-like systems, at metallicities $\lesssim Z_{\mathrm{SMC}}$, should maintain rotational velocities which allow them to exhibit the Be phenomenon from the accretion episode throughout most of their \ac{MS} life (see also section \ref{sec:NBeXB}). 
At higher metallicities, the angular momentum loss through winds and decretion disks may shorten the duration of the Be phenomenon, and so decrease the number of \acp{BeXRB}. 
Moreover, metallicity may  impact mass transfer efficiency (though its impact on the stellar radius, see equations \ref{eq:betaTH} and \ref{eq:th}), and therefore the amount of mass and angular momentum accreted during the mass transfer phase. Given the strong link between the surface rotation and the Be phenomenon, this may in turn impact how many Be stars are formed and for how long they may live. 

\subsection{What can the absence of \acp{BeXRB} with short orbital period imply?}
\label{sec:disc_shortP}
Most of our simulated \ac{SMC} \ac{BeXRB} populations over-predict the number of systems at short orbital periods.  
As mentioned above, this overabundance is closely connected to \ac{US}\acp{SN}. According to our model, \ac{US}\acp{SN} are characterised by low \ac{NS} natal kicks and short separations, and are therefore unlikely to disrupt the binary (assuming $\sigma_{\mathrm{1D}} = 30\ \mathrm{km}\ \mathrm{s}^{-1}$, the binary survival rate is $\sim 98.5\%$).
We can reconcile observations with predictions with two different approaches.
Firstly, we can change our assumptions and so suppress this sub-population.
In the following we list four possible examples:
\begin{itemize}
    \item We could change the stability criteria, allowing only systems with higher mass ratio to experience stable mass transfer (see e.g. $q_{\mathrm{HG}}>0.3$ of figure \ref{fig:stability}), assume a different initial mass ratio distribution or a different stellar response to mass loss.
\item Alternatively we could assign higher natal kick velocities to \acp{NS} born through \ac{US}\acp{SN}. This, however, may contradict observations \citep{Tauris2015,vigna2018formation}. 
\item We could assume a higher accretion efficiency.  As discussed in section \ref{sec:Porb}, the contribution of binaries whose primary has experienced a \ac{US}\ac{SN} decreases when assuming higher accretion efficiency.
\item It is possible that only a fraction of the primaries in \ac{BeXRB}-like systems that engage in case BB mass transfer lose their entire He envelopes and experience \ac{US}\acp{SN} \citep{Tauris2015}. Then only a fraction of these systems should have reduced natal kicks.
Similarly, assuming a different stability criterion for case BB mass transfer may also suppress the population of short-period \ac{BeXRB}-like systems (but see \citealt{vigna2018formation} and \citealt{Romero2020} for the impact on \ac{BNS} observations).
A different treatment of case BB mass transfer and  \ac{US}\acp{SN} could make it possible to explain the orbital period distribution of the observed \acp{BeXRB} with different models of angular momentum carried away during non-conservative mass transfer, which are otherwise ruled out by the predicted peak at short orbital periods. 
These include the StarTrack assumption of $\mathrm{\gamma=1}$ and Jeans mode mass loss (which, however, is strongly disfavoured by other observations, see section \ref{sec:COs}). 
\end{itemize}{}
However, the overall number of predicted \acp{BeXRB} provides an additional constraint (see section \ref{sec:NBeXB}); for example, setting $q_{\mathrm{HG}}>0.3$ without further modifications lowers the overall number of \ac{BeXRB}-like systems below the 69 observations reported in the CK catalogue.

Additional selection effects may also explain the absence of short orbital period \acp{BeXRB} in the observed sample.  \ac{MS} stars in binaries with orbital period below a week may be unable to create a decretion disk because of tidal interactions \citep{Panoglou2016}.  Tidal locking, relevant at periods $\lesssim 10-15$ days, could slow down the rotation of \ac{MS} stars \citep{vdHeuvel1967} and so possibly prevent the formation of a surrounding decretion disk. 

\subsubsection{Kinematics of \ac{BeXRB}-like systems}
Our predictions for kicks and for the number of systems experiencing \ac{US}\acp{SN} could in principle be tested by measuring the space velocities of \acp{BeXRB} in the \ac{SMC}, particularly those with short orbital periods  \citep{Brandt1995,Verbunt1990}.  We expect about half of the systems with period shorter than $\sim 30\mathrm{\, d}$ to have projected velocities below $\sim 20\mathrm{\,km/s}$ (the majority of which due to the low \ac{US}\ac{SN} kicks), and the other half to have projected velocities distributed between 20\,km/s and 100\,km/s (due to the high \ac{CC}\ac{SN} kicks).  

At the $\sim 60$~kpc distance of the \ac{SMC}, 20~km/s corresponds to $\sim70 \mathrm{\mu as/yr}$ of proper motion, while 100~km/s corresponds to $\sim350 \mathrm{\mu as/yr}$.  \acp{BeXRB} in the \ac{SMC} should have end-of-mission proper motion precision of about $\sim 20\mathrm{\mu as/yr}$ with Gaia \citep{2018gdr2};
the proper motions of the fastest \acp{BeXRB} will therefore be detectable, while meaningful upper limits could be placed for the slowest systems. 
Our assumptions on case BB mass transfer and \ac{US}\acp{SN} are thus likely to be testable in the near future with proper motion measurements. 

\subsection{Impact of accretion efficiency on \acp{DCO}}
\label{sec:COs}
Adjusting mass transfer prescriptions to match the Be star mass distribution in \acp{BeXRB} affects the rate of formation \acp{DCO}.  Here we discuss the impact of the various prescriptions on the formation of \acp{DCO} at the metallicity of the \ac{SMC}.
 
According to our default (preferred) model, the formation rate of \acp{BBH} merging within 14 Gyr per unit star forming mass is $\sim 3.5\ (2.1)\times 10^{-5}\,\mathrm{M_{\odot}^{-1}}$. Our default model also yields $\sim 1.5 \times 10^{-5}\,\mathrm{M_{\odot}^{-1}}$ \acp{NSBH} merging within 14 Gyr. This drops by a factor of $\sim 1/3$ for our preferred model. 
In general the contribution of \ac{BeXRB}-like systems to the population of \ac{NSBH} binaries merging within 14 Gyr strongly depends on the prescription for angular momentum loss.
Around $\sim 20\%$ ($\sim 24\%$ for our default model and $\sim 18\%$ for our preferred one) of merging \acp{NSBH} passed through a \ac{BeXRB} phase, if we assume isotropic re-emission of the ejected material from the surface of the accretor.

The  formation rate of \acp{BNS} is sensitive to the assumed accretion efficiency.  It is equal to $\sim 1.7\ (1.2)\times 10^{-5}\,\mathrm{M_\odot^{-1}}$ per unit star forming mass for our default (preferred) model and rises to  $\sim 2.5\times 10^{-5}\,\mathrm{M_\odot^{-1}}$ for fully conservative mass transfer.  The percentage of \acp{BNS} merging within 14 Gyr  varies from $\sim 65\%$ for the default model to $\sim 83\%$ for the preferred model to $\sim 93\%$ when non-accreted material is assumed to carry away the specific angular momentum of a circumbinary ring. A general exception to this are models that assume that non-accreted material leaves the binary from the surface of the donor (Jeans mode), which always predict rates for merging compact binaries 1--2 orders of magnitude lower compared to all other models. Our \ac{BeXRB} study, as well as current gravitational-wave observations, strongly disfavour this mode of angular momentum loss. 
Our models also demonstrate that a significant fraction of the \ac{BNS} population (generally $\gtrsim 1/3$) experienced a \ac{BeXRB}-like phase. 
In particular, in our preferred model, almost all \acp{BNS} ($\sim 96\%$ of the overall population and $\sim 98\%$ among those merging within 14 Gyr) experienced a \ac{BeXRB}-like phase.
These results are consistent with \citet{vigna2018formation}, where the authors study \acp{BNS} in the Galaxy and conclude that the main formation channel begins with stable mass transfer from a \ac{HG} donor onto a \ac{MS} secondary.

For comparison, we briefly describe the number of compact binaries obtained assuming $\gamma = 1$. 
The overall predicted number of \acp{NSBH} per solar mass evolved, is about half the one predicted by our preferred model, while the number of \acp{BBH} remains almost unchanged.
The number of \acp{BNS} is also very similar to our preferred model.  However, the fraction of all types of compact object binaries merging within 14 Gyr is around a factor of two smaller for $\gamma=1$ than for $\gamma = q^{-1}$ (isotropic re-emission).  This is mostly due to decreased binary hardening during the first mass transfer episode when $\gamma=1$, leaving wider binaries behind. 

Quantifying the impact of these variations on the \ac{DCO} merger rates for gravitational-wave observations would require an integration over the cosmic star formation history (see e.g. \citealt{Coen2019}), which is beyond the scope of this paper.

\subsection{On the predicted number of \ac{BeXRB}-like systems}
\label{sec:NBeXB}
The low yield of our simulations suggests that the majority of B stars
in binary systems with a \ac{NS}
are Be stars.  The enhanced ratio of Be stars to B stars in these interacting systems is consistent with the hypothesis that the Be phenomenon originates from accretion, or is boosted by it  \citep{DeMink2013}.
On the other hand, the challenge of producing more \ac{BeXRB}-like systems in simulations suggests that the existing census of \acp{BeXRB} in the \ac{SMC} is already close to complete for orbital periods below a year, assuming negligible selection effects for this orbital period range.

There are, however, several uncertainties/complications which may affect our estimated number of \ac{BeXRB}-like systems.  
\begin{itemize}
	\item The key uncertainty is related to the Be star spindown: for how long will the companion star rotate rapidly enough to remain a Be star in the presence of wind-driven mass loss?  Observations  suggest that Be stars can last for relatively long times at low metallicities; indeed in the \ac{SMC} Be stars have been identified at the end of their \ac{MS} lifetime \citep{Martayan2007SMC}.  Under the simplifying assumption of rigid body rotation, uniform winds from a thin spherical shell carry away a fractional angular momentum 
	\begin{equation}
	\frac{dJ}{J} \sim \frac{2}{3k} \frac{dM}{M},
	\end{equation}
	where $dM/M$ is the fraction of mass lost and $k$ is the radius of gyration \citep{Claret1989,DeMink2013}.  Provided that $\sim 10\ M_\odot$ Be stars lose less than $0.2\ M_\odot$ solar masses through winds, they can remain maximally rotating for their entire \ac{MS} lifetime once spun up.  Estimates of Be star mass loss point to the `weak-wind problem', and indicate that the mass loss should be well within this range \citep{Vink2000,Smith2014,Krticka2014}.    This is consistent with results of detailed studies \citep[e.g. ][]{Heger2000,DeMink2013}, which suggested that   the angular velocity of a rotating $10-20\,\mathrm{M_\odot}$  star naturally evolves toward the Keplerian limit during the \ac{MS} phase. Of course, there are multiple complications to this simple story:
	\begin{itemize}
	\item How efficiently is angular momentum transported through the star, and how close is the star to rigid rotation \citep{packet1981spin,Zahn1992,Spruit2002,Petrovic2005_b}?  
	\item How does the moment of inertia of the \ac{MS} star change as it evolves?
	\item How much mass and angular momentum is lost through the decretion disk? 
	The decretion disk may be a key regulator of the star's angular momentum  \citep{Rivinius2013}: even if only a small fraction of the total mass loss is through the disk, the $\propto \sqrt{r}$ scaling of Keplerian angular momentum with radius ensures that the disk can make a significant contribution to the angular momentum loss.  
	Recent estimates \citep{Rimulo2018} indicate that the typical angular momentum loss through the decretion disk may allow $\sim10\,\mathrm{M_\odot}$ stars at the \ac{SMC} metallicity to maintain angular frequencies at a significant fraction of critical. 
	However, the amount of mass loss through the disk has so far only been measured for the Galactic Be star $\mathrm{\omega CMa}$ \citep{Carciofi2012}, and the uncertain disk dynamics, density, and stable size make it very challenging to model the amount of angular momentum lost through the disk \citep{Okazaki2005,Labadie_Bartz_2017,Rimulo2018,Brown2019}.
	    \item What is the threshold rotational frequency for a B star to appear as a Be star \citep{Rivinius2013}?  \citet{Huang2010} find that this threshold may vary from $0.93$ of critical rotation for $\lesssim 4\, \mathrm{M_\odot}$ stars to $0.56$ of critical for $\gtrsim  8.6\, \mathrm{M_\odot}$ stars.
\end{itemize}
    \item Stellar rotation also impacts the total \ac{MS} lifetime, and hence the duration of the \ac{BeXRB} phase and the number of \ac{BeXRB}-like systems.  In particular, COMPAS models do not account for the uncertain amount of rotational mixing and the consequent additional hydrogen available for nuclear burning in rapidly rotating stars \citep{Spruit2002,Langer2008}. 

    \item Another uncertainty, which might affect the predicted numbers of simulated \acp{BeXRB}, is the magnitude of \ac{SN} kicks.  There is ongoing debate about the frequency of low natal kicks in the observed pulsar population \citep[e.g.,][]{Hobbs:2005, Arzoumanian2002, IgoshevVerbunt:2017,Bray2016,Bray2018,Scheck2004,Scheck2006,Nordhaus2010,Nordhaus2012,Muller2017,Janka2017,Gessner2018,2020Powell}. For example, \citet{Podsiadlowski2004} argued that low-mass iron core collapse \acp{SN} might experience lower kicks compared to typical \ac{CC}\acp{SN}.
    \item As shown in section \ref{sec:Results}, the stability criteria for mass transfer may play a key role. Some of the mass transfer episodes designated as dynamically unstable in our simulations could, in fact, be stable and produce \acp{BeXRB} \citep{Pavlovskii2017}.  However, increasing the number of systems undergoing stable $\mathrm{HG\rightarrow MS}$ mass transfer with low mass ratio $q$ would considerably increase the already overestimated population of \ac{BeXRB}-like systems at short orbital period, assuming the primaries experience low-kick \ac{US}\acp{SN} which allow such binaries to survive.
    \item Conversely, some of the systems undergoing a \ac{CE} event before the first \ac{SN} 
    may contribute to the observed population of \acp{BeXRB} if:
    \begin{itemize}
        \item a significant amount of mass is accreted during dynamically unstable mass transfer, which appears unlikely \citep{MacLeodRamirezRuiz:2015,De:2019};        
        \item not much mass needs to be accreted to spin up the \ac{MS} stars \citep{packet1981spin} (or the Be phenomenon is not linked to accretion), so \ac{CE} events are sufficient to produce Be stars --- however, to avoid \acp{BeXRB} with low-mass companions that are not present in the observational sample, \ac{CE} events with low-mass companions would lead to mergers; or
        \item the systems which experience a subsequent dynamically stable mass transfer episode after a \ac{CE} event acquire a significant amount of mass, despite the short duration of this mass transfer episode and the significant difference in the thermal timescales of the donor and the accretor.
    \end{itemize}
    \item The number of simulated \ac{SMC} \acp{BeXRB} could be slightly increased by considering the contribution of post-\ac{MS} stars to the Be star population. 
    Indeed the CK lists luminosity classifications that range between type II and type V. 
    However, we only consider \ac{MS} stars as the luminosity classification may not be indicative of the actual photometric magnitudes (see e.g. \citealt{McBride2008}), due to both difficulties of the measurements involved and the heterogeneity of the adopted classification methods.
    \item 
    The significant scatter of the Corbet relation could shift  some of the 25 \acp{BeXRB} with unknown orbital properties in the CK catalogue to orbital periods longer than the maximum measured one of $\sim 520$ days. This seems reasonable given some of their relatively large spin periods (3 have spin periods above the maximum of the 44 systems with known orbital periods and 9 have it above the one corresponding to $P_{\mathrm{orb}} = 520$ days according to the fit of equation (\ref{eq:Corbet})).  This would reduce the number of observed systems with $P_{\mathrm{orb}} < 520$ days that we are trying to match. 
    This change would not be sufficient to significantly impact our main conclusions, including the choice of our preferred model. 
    Assuming that only 44 \acp{BeXRB} are actually observed with orbital period $\lesssim520\,$d (see below) would however suggest that the survey is not complete or 
    weaken our finding that most of the 
    B stars that accreted a considerable amount of mass must exhibit the Be phenomenon for the majority of their \ac{MS} lives. 
\end{itemize}
Finally we should also consider the significant uncertainties in the total number of observed \acp{BeXRB} in the \ac{SMC}. 
\citet{Maravelias2019} and \citet{Haberl&Sturm2015} argue that the \ac{SMC} contains about $\sim 120$ \acp{HMXRB}, of which almost all are supposed to contain a Be star. Only about half of those (the 69 reported in the CK catalogue) show X-ray pulsations, clearly identifying \acp{NS} as the accreting compact object in the binary. The remaining systems may contain \acp{BH}, white dwarfs, or \acp{NS} with unknown but most likely large ($\gtrsim 100\,$s, \citealt{Haberl&Sturm2015}) spin periods.
Further observations are needed to measure the fraction of \acp{HMXRB} not exhibiting X-ray pulsations that contain \acp{NS}. If this fraction is high, or if an appreciable population of $\mathrm{NS+MS}$ systems with no X-ray emission is discovered, our models may need further revision.

\subsection{Observational constraints}
Further, more detailed observations of \acp{BeXRB} in the \ac{SMC} will provide the strongest tests of our findings.  As mentioned in previous sections, (i) the mass distribution of Be stars in \acp{BeXRB}, (ii) a more complete set of measurements of their orbital period and eccentricity, (iii) the detections of radio pulsars in wide binary systems with Be stars, (iv) the classification of the compact objects in the remaining \acp{HMXRB} of the \ac{SMC} and (v) the measurements of \ac{BeXRB} space velocities will provide crucial constraints on our binary evolution models. 

Observations of other stages of binary evolution can provide additional constraints on the uncertain evolutionary phases parameterised in population synthesis models \citep[e.g.,][]{Eldridge2017}.  These include observations of compact-object binaries and inference on gravitational-wave source populations \citep[e.g.,][]{Barrett2018}.  They also include observations of earlier evolutionary stages, particularly \ac{MS}-\ac{MS} mergers and \ac{MS}+helium star binaries.  However, models of the frequency of merger products, such as blue stragglers \citep[e.g.,][]{DeMink2013,Bodensteiner2020} as well as possible merger transients, such as luminous red novae \citep[e.g.,][]{Howitt:2020}, require further assumptions about the uncertain physics of merger dynamics, the treatment of contact systems, the possibility of chemically homogeneous evolution, etc.  The impact of mass transfer efficiency $\beta$ could generally be explored by observing the mass gain and rejuvenation of binary companions \citep{vdHeuvel1967, Schneider2015}.  

Binaries consisting of \ac{MS} stars and stripped helium companions have a more immediate connection to \acp{BeXRB}, though the primaries in most such systems will not collapse into \acp{NS} or disrupt the binary by \ac{SN} kicks.  In our preferred model, we find that one or two thousand non-mass transferring \ac{MS}+He star binaries should be present in the \ac{SMC} with $\gtrsim 3\ M_\odot$ \ac{MS} companions.  However, the uncertainty on this number is at least as large as the relative uncertainty on the number of \acp{BeXRB} discussed in section \ref{sec:NBeXB}.  In addition to modelling uncertainties, we must deal with challenging selection effects, including possible misclassifications (see, e.g., the very different proposed nature for the unseen companion in the LB-1 system \citealt{Liu2019,Eldridge:2020,shenar2020hidden}).

\section{Conclusions}
\label{sec:Conclusions}
Our models strongly suggest that observed \acp{BeXRB} experienced one or more dynamically stable mass transfer episodes initiated by the progenitor of the \ac{NS}.

We find that stable mass transfer with accretion efficiency ${{\gtrsim} 0.3}$ matches the observed properties of \acp{BeXRB} in the \ac{SMC}. The finding highlights the utility of rapid population synthesis tools with easily modified recipes for exploring the impact of a large number of alternative models, including models with ad hoc constraints.
A similar result was presented by \citet{presentation}. 
This points to an increased incidence of Be stars in interacting binaries, possibly due to spin-up generated by mass accretion.  

We show that a deeper understanding of selection effects and of the occurrence rate of Be vs B stars in binary systems with a \ac{NS} can improve inference on the critical mass ratio for stable mass transfer, accretion efficiency and angular momentum loss during $\mathrm{HG\rightarrow MS}$ mass transfer and/or \ac{US}\ac{SN} kicks.

Our simulations also suggest that currently observed \acp{BeXRB} were born when the \ac{SMC} \ac{SFR} peaked, at least partially explaining the observed abundance of \acp{BeXRB}.
This is supported by the predicted age of interacting systems at the \ac{NS}+\ac{MS}-star stage.
In terms of binary evolution, metallicity does not appear to play a major role in explaining the abundance of \acp{BeXRB} in the \ac{SMC}. 
Our models also suggest that metallicity may not be crucial even in terms of Be vs B star occurrence in \acp{BeXRB}. We find that there is not much room for a population of currently non-interacting $\mathrm{NS+MS}$ systems (which could represent the fraction of $\mathrm{NS+MS}$ binaries containing a normal B star): the sample of observed \acp{BeXRB} in the \ac{SMC} must be close to  complete, at least for relatively close systems. Only then can our models produce enough \acp{BeXRB} to be consistent with observations.

The requirements placed on mass transfer stability and accretion efficiency by \ac{BeXRB} observations impact models of other massive binaries, including predictions for \ac{DCO} mergers.

\section*{Acknowledgements}
The authors thank the anonymous referee, K.~Auchettl, A.~Carciofi, M.J.~Coe, J.J.~Eldridge, M.~Gilfanov, R.~Hirai, C.~Knigge, M.~MacLeod, E.~Ramirez-Ruiz, M.~Renzo and T.~Rivinius for helpful discussions. 
We also thank V.~Antoniou, M.~Heida and S.~Motta for useful information on X-ray observations and their limitations.
A.V-G. acknowledges funding support from Consejo Nacional de Ciencia y Tecnolog\'ia (CONACYT) and support by the Danish National Research Foundation (DNRF132).
IM is a recipient of the Australian Research Council Future Fellowship FT190100574.   IM and PP acknowledge that  part of this work was performed at the KITP, which is supported in part by the National Science Foundation under Grant No.~NSF PHY-1748958, and at the Aspen Center for Physics, which is supported by National Science Foundation grant PHY-1607611.
MN is supported by a Royal Astronomical Society Research Fellowship.   

\section*{Data availability}
Simulations in this paper made use of the COMPAS rapid population synthesis code which is freely available at \url{http://github.com/TeamCOMPAS/COMPAS}. 
The data underlying this article will be shared on request to the corresponding author.


\bibliographystyle{mnras} 
\bibliography{bib2}
    
\label{lastpage}
\end{document}